\documentclass[11pt,epsf]{article}
\usepackage{amsmath}
\usepackage{graphicx}
\graphicspath{{./figs}}

\usepackage[merge,numbers,compress]{natbib}
\usepackage[T1]{fontenc}
\usepackage{booktabs}
\usepackage{xcolor}
\usepackage{xspace}
\usepackage{subfigure}
\usepackage{dcolumn}
\usepackage{placeins}
\usepackage{amsfonts}
\usepackage{yhmath}
\usepackage[colorlinks=true,citecolor=blue!95!black,linkcolor=black]{hyperref}
\usepackage{caption}
\usepackage[subrefformat=parens,position=top,skip=-15pt,margin=15pt,justification=justified,singlelinecheck=false]
 {subcaption}
\usepackage[normalem]{ulem}
\usepackage{float}
 \usepackage{verbatim}
\usepackage{siunitx}
\usepackage{geometry}
\usepackage{pdflscape}
\usepackage{ifthen}
\usepackage[capitalize]{cleveref}
\makeatletter
\AddToHook{cmd/appendix/before}{\def\cref@section@alias{appendix}\def\cref@subsection@alias{appendix}}
\makeatother

\usepackage{tikz}
\usetikzlibrary{arrows.meta,calc}
\usepackage{xparse}
\usepackage[disable]{todonotes}  
\usepackage{multirow}

\usepackage{listofitems} 

\definecolor{red}{HTML}{FF00aa}
\definecolor{black}{HTML}{000000}   
\definecolor{gray}{HTML}{808080}    
\definecolor{red}{HTML}{EB3323}     
\definecolor{yellow}{HTML}{D2A641}  
\definecolor{green}{HTML}{377D22}   
\definecolor{blue}{HTML}{001EF5}    
\definecolor{pink}{HTML}{EB46F8}    

\usepackage{slashed}
\usepackage{amsmath}
\usepackage{amsfonts}
\usepackage{amssymb}

\usepackage{tabularray}
\UseTblrLibrary{amsmath}

\def\citere#1{\mbox{Ref.~\cite{#1}}}
\def\citeres#1{\mbox{Refs.~\cite{#1}}}

\newcommand{\newc}{\newcommand}
\newc{\beq}{\begin{equation}}
\newc{\eeq}{\end{equation}}
\newc{\beqn}{\begin{eqnarray}}
\newc{\eeqn}{\end{eqnarray}}
\newc{\bit}{\begin{itemize}}
\newc{\eit}{\end{itemize}}
\newc{\ben}{\begin{enumerate}}
\newc{\een}{\end{enumerate}}
\newc{\bce}{\begin{center}}
\newc{\ece}{\end{center}}
\newc{\bfi}{\begin{figure}}
\newc{\efi}{\end{figure}}

\newcommand{\rd}{\mathrm d}

\newcommand{\rT}{{\mathrm{T}}}

\newcommand{\rL}{{\mathrm{L}}}

\newcommand{\ie}{\emph{i.e.}\ }

\newcommand{\GeV}{\ensuremath{\,\text{GeV}}\xspace}
\newcommand{\TeV}{\ensuremath{\,\text{TeV}}\xspace}
\newcommand{\fb}{{\ensuremath\unskip\,\text{fb}}\xspace}

\newcommand{\Pj}{\ensuremath{\text{j}}\xspace}
\newcommand{\Pp}{\ensuremath{\text{p}}}
\newcommand{\Pe}{\ensuremath{\text{e}}\xspace}

\newcommand{\PW}{\ensuremath{\text{W}}\xspace}
\newcommand{\PZ}{\ensuremath{\text{Z}}\xspace}

\newcommand{\MW}{\ensuremath{M_\PW}\xspace}

\newcommand{\MZ}{\ensuremath{M_\PZ}\xspace}

\newcommand{\GZ}{\ensuremath{\Gamma_\PZ}\xspace}

\newcommand{\GW}{\ensuremath{\Gamma_\PW}\xspace}

\newcommand{\GF}{\ensuremath{G_\mu}}
\newcommand{\alphas}{\ensuremath{\alpha_\text{s}}\xspace}

\newcolumntype{.}{D{.}{.}{-1}}
\newcolumntype{d}[1]{D{.}{.}{#1}}
\colorlet{tableoverheadcolor}{gray!37.5}
\colorlet{tableheadcolor}{gray!25}
\colorlet{tablerowcolor}{gray!12.5}

\def\draftdate{\relax}
\def\mda{\relax}
\def\mua{\relax}
\def\mla{\relax}
\def\draft{
\def\thtystars{******************************}
\def\sixtystars{\thtystars\thtystars}
\typeout{}
\typeout{\sixtystars**}
\typeout{* Draft mode!
         For final version remove \protect\draft\space in source file *}
\typeout{\sixtystars**}
\typeout{}
\def\draftdate{\today}
\def\mua{\marginpar[\boldmath\hfil$\uparrow$]%
                   {\boldmath$\uparrow$\hfil}\color{black}%
                    \typeout{marginpar: $\uparrow$}\ignorespaces}
\def\mda{\color{red}\marginpar[\boldmath\hfil$\downarrow$]%
                   {\boldmath$\downarrow$\hfil}%
                    \typeout{marginpar: $\downarrow$}\ignorespaces}
\def\mla{\marginpar[\boldmath\hfil$\rightarrow$]%
                   {\boldmath$\leftarrow $\hfil}%
                    \typeout{marginpar: $\leftrightarrow$}\ignorespaces}
\def\Mua{\marginpar[\boldmath\hfil$\Uparrow$]%
                   {\boldmath$\Uparrow$\hfil}\color{black}%
                    \typeout{marginpar: $\uparrow$}\ignorespaces}
\def\Mda{\color{red}\marginpar[\boldmath\hfil$\Downarrow$]%
                   {\boldmath$\Downarrow$\hfil}%
                    \typeout{marginpar: $\downarrow$}\ignorespaces}
\def\Mla{\marginpar[\boldmath\hfil\textcolor{red}{$\Rightarrow$}]%
                   {\boldmath\textcolor{red}{$\Leftarrow $}\hfil}%
                    \typeout{marginpar: $\leftrightarrow$}\ignorespaces}
\overfullrule 5pt
\oddsidemargin 15mm
\marginparwidth 29mm
}

\newcommand{\mc}{\mathcal}

\newcommand{\pt}[1]{p_{\rT,{#1}}}
\newcommand{\nnb}{\nonumber}

\newcommand*{\relu}{\text{ReLU}}
\newcommand{\rll}{\ensuremath{r_{\rL \rL}}}
\newcommand{\rlltilde}{\ensuremath{\tilde{r}_{\rL \rL}}}

\newcommand*{\noun}[1]{\textsc{#1}}  
\newcommand*{\POWHEG}{\noun{Powheg}}
\newcommand*{\POWHEGBOX}{\noun{Powheg-Box}}
\newcommand*{\POWHEGBOXRES}{\noun{Powheg-Box-Res}}

\newcommand*{\recolatwo}{\noun{Recola}~2}

\newcommand*{\autoencoder}{\text{AE}}
\newcommand*{\ffnn}{\text{FFNN}}
\newcommand*{\rfr}{\text{RFR}}
\newcommand*{\pn}{\text{PN}}

\newcommand*{\true}{\text{truth}}

\newcommand*{\lo}{\text{LO}}
\newcommand*{\nlo}{\text{NLO}}
\newcommand*{\lows}{\text{LO}\ensuremath{+}\text{Sud.}}

\newcommand*{\lops}{\text{LO}\ensuremath{+}\text{PS}}
\newcommand*{\nlops}{\text{NLO}\ensuremath{+}\text{PS}}

\DeclareSIUnit{\gev}{\giga \electronvolt}
\DeclareSIUnit{\barn}{b}

\begin{document}

\title{\hfill ~\\[-30mm]
\phantom{h}\hfill\mbox{\small COMETA-2026-20, FR-PHENO-2026-010, LAPTH-036/26, MPP-2026-122}\\[1cm]
\vspace{13mm}\textbf{
Higher-order effects in amplitude-assisted polarisation extraction with machine-learning techniques}}
\date{} 
\author{
Juan M. Cruz-Martinez$^{\,1\,}$\footnote{\texttt{jcruz@us.es}},
Jakob Linder$^{\,2,3\,}$\footnote{\texttt{linder@mpp.mpg.de}}, 
Mathieu Pellen$^{\,4\,}$\footnote{\texttt{mathieu.pellen@physik.uni-freiburg.de}},\\
Giovanni Pelliccioli$^{\,5,6\,}$\footnote{ \texttt{giovanni.pelliccioli@unimib.it}}, 
Emanuele Re$^{\,5,6\,}${\footnote{\texttt{emanuele.re@mib.infn.it}}\,\,\,\footnote{On leave of absence from \emph{Laboratoire d'Annecy de Physique Th\'eorique (LAPTh), CNRS, USMB, F-74940 Annecy, France.}}}
\\[9mm]
{\small \it $^1$ Departamento de Fisica Atomica, Molecular y Nuclear, Facultad de Fisica, Universidad de Sevilla}\\
{\small \it E-41080 Sevilla, Spain}\\[3mm]
{\small\it $^2$ Max-Planck-Institut f\"ur Physik,}
{\small \it Boltzmannstrasse 8, 85748 Garching, Germany}\\[3mm]
{\small\it $^3$ Technische Universit\"at M\"unchen,}
{\small \it James-Franck-Strasse 1, 85748 Garching, Germany}\\[3mm]
{\small\it $^4$ Albert-Ludwigs Universit\"at Freiburg, Physikalisches Institut}\\
{\small \it Hermann-Herder-Stra{\ss}e 3, D-79104, Freiburg im Breisgau, Germany}\\[3mm]
{\small\it $^5$ Dipartimento di Fisica, Universit\`a degli Studi di Milano-Bicocca,}
{\small \it Piazza della Scienza 3, 20126 Milano, Italy}\\[3mm]
{\small\it $^6$ Istituto Nazionale di Fisica Nucleare, Sezione di Milano-Bicocca,}
{\small \it Piazza della Scienza 3, 20126 Milano, Italy}\\[3mm]
}

\maketitle

\begin{abstract}
\noindent
With increasing experimental precision, the prospect of extracting the polarisation of electroweak gauge bosons is becoming particularly attractive.
To this end, regression and classification procedures based on precise and accurate theoretical predictions are becoming increasingly important. In this work, we present the first amplitude-assisted regression procedure at next-to-leading-order accuracy in QCD, supplemented with parton-shower effects, using machine-learning techniques to extract the rate of longitudinal-boson production in high-energy collisions.
Several neural-network architectures are presented and benchmarked against a standard random-forest regressor, demonstrating the robustness of the results for di-boson production at the LHC.
\end{abstract}
\thispagestyle{empty}
\vfill
\newpage
\setcounter{page}{1}

\hrule
\tableofcontents
\hspace*{2cm}\hrule

\section{Introduction} \label{sec:intro}
Studying the polarisation state of electroweak (EW) gauge bosons at TeV-scale colliders provides a powerful probe of the electroweak symmetry breaking (EWSB) pattern realised in nature, presently described within the Standard Model (SM) by the Higgs mechanism. Through the EWSB, $\PW$ and $\PZ$ bosons acquire a non-vanishing mass and a longitudinal degree of freedom (on top of the two transverse ones).
Consequently, observing any anomaly in the production rates or kinematic features of longitudinally polarised bosons in collider processes would signal the presence of underlying dynamics not described by the SM, and more specifically, a different realisation of the EWSB mechanism.

In light of polarisation-focused Run-2 and Run-3 LHC measurements \cite{Aaboud:2019gxl,CMS:2020etf,CMS:2021icx,ATLAS:2022oge,ATLAS:2023zrv,ATLAS:2024qbd,ATLAS:2025wuw}, the theory community has put a lot of effort into advancing the understanding, the automation, and the precision of the simulations of polarised-boson signals
\cite{Ballestrero:2017bxn,BuarqueFranzosi:2019boy,Ballestrero:2019qoy,Ballestrero:2020qgv,Denner:2020bcz,Denner:2020eck,Poncelet:2021jmj,Denner:2021csi,Le:2022lrp,Le:2022ppa,Denner:2022riz,Pelliccioli:2023zpd,Denner:2023ehn,Dao:2023kwc,Grossi:2024jae,Denner:2024tlu,Dao:2024ffg,Carrivale:2025mjy,Haisch:2025jqr,DelGratta:2025xjp,Pelliccioli:2025com,Denner:2025xdz,Basu:2025zds,Le:2026oux,Pellen:2021vpi,Pellen:2022fom,Hoppe:2023uux,Javurkova:2024bwa}, in order to provide ATLAS and CMS with \emph{polarised templates} for data fitting.
The strength of this approach is that, once background contributions and genuine experimental effects are subtracted, the signal events are further analysed using separate templates for specific polarisation states of the intermediate bosons. This enables the determination of polarisation fractions from differential distributions of experimental data, incorporating in a consistent manner interference terms and spin correlations.

Alternatively, it has been proposed to promote the ratios of specific matrix elements to physical observables through the use of Neural Networks~\cite{Grossi:2023fqq}.
As opposed to other approaches in the literature, the method has the advantage of ensuring that all information available in the data is used for the extraction of gauge-boson polarisation.
In addition, the interpretation or extraction can be done at the event level as opposed to the integrated level, implying that it can be used as a polarisation tagger.
Also, the method is particularly efficient numerically as it only requires the evaluation of the ratio of matrix elements at each event.
In Ref.~\cite{Grossi:2023fqq}, the proof of concept has been demonstrated for the tagging of the longitudinal polarisation of a Z boson produced in association with a jet at the LHC
at leading order (\lo{}) accuracy supplemented with parton-shower corrections.

Given the current and upcoming precision at the LHC, precise theoretical predictions are not only needed in theory-experiment comparison but also in the extraction of fundamental quantities.
This implies that it is necessary to extend the method of  Ref.~\cite{Grossi:2023fqq} to next-to-leading-order (\nlo{}) accuracy, possibly matched with parton shower effects. This is the main goal of the current work.

On the technical aspect of the machine-learning approaches, the extraction of polarisations of EW bosons in LHC processes has been tackled in several ways: deep Neural Networks (NNs) \cite{Searcy:2015apa,Lee:2018xtt,Lee:2019nhm,Murphy:2019utt,Grossi:2020orx}, 
random forests and gradient-boosted decision trees \cite{Murphy:2019utt,Dey:2021sug}, generative adversarial networks \cite{Li:2021cbp}, convolutional NNs \cite{Kim:2021gtv}, 
wide NNs \cite{Grossi:2023fqq}, physics-informed reinforcement-learning \cite{Dillon:2025dxr}. 
On the experimental side, all polarisation-oriented ATLAS and CMS analyses \cite{Aaboud:2019gxl,CMS:2020etf,CMS:2021icx,ATLAS:2022oge,ATLAS:2023zrv, ATLAS:2024qbd,ATLAS:2025wuw} rely on ML models, which are typically trained with events simulated at \lo{} accuracy (possibly with multi-jet merging). The impact of higher-order QCD corrections on kinematic-based NN models (trained with \lo{}-accurate events) has been studied in the context of the recent ATLAS analysis, which led to the measurement of joint polarisation fractions in $\PW\PZ$ events \cite{ATLAS:2022oge}. However, a study of the impact of higher-order corrections on the training and testing of ML models which target polarisation tagging does not exist to date.
The present work fills this gap by developing a general approach with \nlo{} QCD (matched to a parton shower) accuracy.
In particular, as a showcase example, we study the di-boson process with two Z bosons decaying leptonically, and present several strategies to build NNs for this purpose.

An extension beyond \nlo{} QCD would require the generation of polarised-boson events at NNLO QCD and \nlo{} EW accuracy matched to parton/photon shower. While within reach using similar strategies as for off-shell di-boson production \cite{Lombardi:2021rvg,Chiesa:2020ttl,Lindert:2022qdd}, this is technically involved and not supported by any generator for now.  However, we are confident that our results at \nlo{} QCD capture the bulk of the relevant effects in polarisation tagging, therefore allowing for extensions to higher perturbative orders without conceptual bottlenecks.

The structure of this article reads as follows:
In \cref{sec:strategy}, the general strategy is discussed, in particular, how the method proposed in Ref.~\cite{Grossi:2023fqq} has to be modified to account for higher-order corrections in QCD.
\Cref{sec:ml} focuses specifically on various machine-learning approaches for tagging polarisation in a fully realistic experimental setup.
In \cref{sec:results} we show the results of our study.
Finally, \cref{sec:conclusion} contains concluding remarks.
\cref{app:more_distrib,app:rfr_diff_inputs}, display additional kinematic plots for the different ML models and investigate the effect of using different input choices in the training of them.
More details on the training of the models are given in \cref{app:hyperparameters}.

\section{Polarisation tagging and QCD higher-order corrections} 
\label{sec:strategy}

\subsection{General idea}

The amplitude of a process with two resonant gauge bosons decaying into lepton pairs can be written as follows (in the unitary gauge):
\begin{equation}
\label{eq:Mlep}
\mathcal{M} = \mathcal{M}^{\mc P}_{\mu\rho} \left(-g^{\mu\nu}+\frac{k_1^{\mu}  k^{\nu}_1}{M_{V_1}^2}\right) \mathcal{M}^{\mc D_1}_{\nu} \left(-g^{\rho\eta}+\frac{k_2^{\rho}  k^{\eta}_2}{M_{V_2}^2}\right)  \mathcal{M}^{\mc D_2}_{\eta} \prod_{i=1}^2 \frac{\rm i}{k_i^2 - M_{V_i}^2 + {\rm i} \Gamma_{V_i} M_{V_i}},
\end{equation}
where $\mc{M}^{\mc P}$, $\mc{M}^{\mc D_1}$, and $\mc{M}^{\mc D_2}$ describe the production and the two decay parts of the amplitude, respectively.
The mass and the width of the resonances are denoted by $M_{V_i}$ and $\Gamma_{V_i}$.
Furthermore, the tensor part of the propagators takes the form
\begin{equation}
-g^{\mu\nu} + \frac{k_i^{\mu}k_i^{\nu}}{M_{V_i}^2} = \sum_{\lambda = 1}^4 \varepsilon^{\mu}_\lambda(k_i)
\varepsilon^{\nu^*}_{\lambda}(k_i)\,\,,
\end{equation}
where $\{\varepsilon^{\mu}_\lambda(k)\}$ denote the polarisation vectors of the massive gauge bosons.
The sum runs over four polarisation states: three physical states and an unphysical one.
In this article, the labels $\rL$, $+$, and $-$ are used for the longitudinal, right-handed, and left-handed states, respectively.
The fourth, unphysical degree of freedom cancels against Goldstone-boson contributions at any order in perturbation theory \cite{Denner:2021csi}.
When the right- and left-handed polarisations are summed coherently, \ie including interference between them, we denote the contribution with $\rT$, which stands for transverse polarisation.

The (unpolarised) amplitude, including both production and decay, reads
\begin{equation}\label{eq:Msum}
\mathcal{M} = \sum_{\lambda_1, \lambda_2} \mathcal{M}_{\lambda_1,\lambda_2}\,,\qquad \lambda_1, \lambda_2=\rL, +,- \, ,
\end{equation}
where $\mathcal{M}_{\lambda_1,\lambda_2}$ is the amplitude with two polarised intermediate gauge bosons, defined as:
\begin{equation}\label{eq:Mlep_pol_lambda}
\mathcal{M}_{\lambda_{1} \lambda_{2}} = \big[\mathcal{M}^{\mc P}_{\mu \rho} \varepsilon^{\mu}_{\lambda_{1}}(k_1) \varepsilon^{\rho}_{\lambda_{2}}(k_2) \big]
\big[\varepsilon^{*\,\nu}_{\lambda_{1}}(k_1)\mathcal{M}^{\mc D_1}_{\nu} \big]
\big[\varepsilon^{*\,\eta}_{\lambda_{2}}(k_2)\mathcal{M}^{\mc D_2}_{\eta} \big]
\prod_{i=1}^2 \frac{i}{k_i^2 - M_{V_i}^2 + i \Gamma_{V_i} M_{V_i}}\,.
\end{equation}
Squaring the unpolarised amplitude leads to
\begin{equation}
\label{eq:interfpol}
\left|\mathcal{M}\right|^2 = {\sum_{\lambda_{1}^{\phantom{\prime}}, \lambda_{2}^{\phantom{\prime}}}\left|
  \mathcal{M}_{\lambda_{1}^{\phantom{\prime}} \lambda_{2}^{\phantom{\prime}}}\right|^2} + {\sum_{\left\{\lambda_{1}^{\phantom{\prime}},\lambda_{2}^{\phantom{\prime}}\right\} \neq \left\{\lambda_{1}^{\prime},\lambda_{2}^{\prime}\right\}}
  \mathcal{M}_{\lambda_{1}^{\phantom{\prime}} \lambda_{2}^{\phantom{\prime}}}^{ *} \mathcal{M}_{\lambda_{1}^{\prime} \lambda_{2}^{\prime}}}\,,\qquad \lambda_1,\lambda_2=\rL, \pm\,,
\end{equation}
with the first sum enclosing the incoherent sum over polarised squared amplitudes and the second featuring all interference terms.
Hence, the transversely ($\rT$) polarised amplitude reads
\begin{equation}
  \left|\mathcal{M}_{\rT\lambda}\right|^2 =
  \left|\mathcal{M}_{+\lambda}\right|^2 + \left|\mathcal{M}_{-\lambda}\right|^2
  +
  2\,{\rm Re}\,\left(\mathcal{M}_{+\lambda}^{ *}\mathcal{M}^{\phantom{*}}_{-\lambda}\right)\,,
\end{equation}
with $\lambda$ arbitrary, leading to
\begin{equation}\
  \label{eq:L+T}
  \left|\mathcal{M}\right|^2 =
  \left|\mathcal{M}_{\rL\rL}\right|^2 + \left|\mathcal{M}_{\rT\rT}\right|^2
  +\left|\mathcal{M}_{\rL\rT}\right|^2
  +\left|\mathcal{M}_{\rT\rL}\right|^2
  + (\textrm{interference terms})\,.
\end{equation}
The term $\left|\mathcal{M}_{\rL\rL}\right|^2$ denotes the doubly longitudinally polarised squared amplitude, which is the one that we are most interested in, in this work.
It is worth mentioning that polarisations are well-defined only for on-shell gauge bosons.
This implies that theoretical predictions with polarised gauge bosons do only account for resonant contributions, while non-resonant contributions are regarded as a background \cite{Denner:2020bcz}.
A particularly well-suited approximation in this context is the double-pole approximation (DPA), which has been widely used in the context of polarised-boson theory templates
\cite{Ballestrero:2017bxn,BuarqueFranzosi:2019boy,Ballestrero:2019qoy,Ballestrero:2020qgv,Denner:2020bcz,Denner:2020eck,Poncelet:2021jmj,Denner:2021csi,Le:2022lrp,Le:2022ppa,Denner:2022riz,Pelliccioli:2023zpd,Denner:2023ehn,Dao:2023kwc,Grossi:2024jae,Denner:2024tlu,Dao:2024ffg,Carrivale:2025mjy,Haisch:2025jqr,DelGratta:2025xjp,Pelliccioli:2025com,Denner:2025xdz,Basu:2025zds,Le:2026oux} and will also be used for benchmark simulations in the present work.
We stress that since we are not interested in discriminating between left- and right-handed polarisation states, we will only consider the longitudinal ($\rL$) and transverse ($\rT$) states, with the latter being defined as the coherent sum of left- and right-handed polarisation states.

The fully differential unpolarised and doubly-longitudinal cross sections schematically read
\begin{align}\label{eq:diffcross}
{\rm d} \sigma_{\rm unp} = \frac{1}{F} \,\left|\mathcal{M}\right|^2 \,{\rm d} \Phi,\qquad 
{\rm d} \sigma_{\rL\rL} = \frac{1}{F} \, \left|\mathcal{M}_{\rL\rL}\right|^2 \, {\rm d} \Phi\,,
\end{align}
where the flux factor and phase-space measure are denoted by $F$ and ${\rm d} \Phi$, respectively.
The differential longitudinal fraction in a generic observable $\mathcal{O}$ is hence defined as
\begin{align}
\label{eq:fraction}
 f_{\rL\rL}(\mathcal{O})= \frac{{\rm d} \sigma_{\rL\rL}}{\rm d \mathcal{O}}\bigg/\frac{\rm d \sigma_{\rm unp}}{\rm d \mathcal{O}} \,.
\end{align}
Comparing \cref{eq:fraction} with \cref{eq:diffcross} makes it transparent that the fully differential polarisation fraction
is a simple ratio of the longitudinal matrix element over the unpolarised one. Using the same notation as in \citere{Grossi:2023fqq}, we define such a ratio as
\begin{align}
\label{eq:rLL}
 \rll = \frac{\left|\mathcal{M}_{\rL\rL} \right|^2}{\left|\mathcal{M}\right|^2} .
\end{align}
With the full knowledge of the momenta, \rll{} can be used as a reweighting factor to obtain polarised samples from unpolarised Monte Carlo (MC) predictions.
Instead, if only physical outputs are available, one needs to regress the function \rll{} without the full knowledge of the momenta entering it.

Reference~\cite{Grossi:2023fqq} promoted the idea of obtaining predictions with longitudinally polarised gauge bosons from predictions with unpolarised bosons upon using a reweighting factor which boiled down to a ratio of the matrix element with longitudinally polarised bosons over the one with unpolarised bosons, dubbed $r_\rL$.
The method is exact at leading order (\lo{}) and was applied to \lo{} predictions matched to parton-shower (PS) corrections for the production of a Z boson in association with a jet at the LHC.

While this method provides an efficient reweighting strategy, it does not allow for the tagging of the longitudinal bosons in an experimental environment.
Indeed, the ratio $r_\rL$ is not a physical observable as it relies on the knowledge of the initial four-momenta and the partonic channel.
To overcome this limitation, a NN was introduced to obtain an approximate version of the ratio, dubbed $\tilde r_\rL$, which can be evaluated for each event based on the experimentally accessible observables (the visible final-state momenta).
Given that the ratio typically takes values between 0 and 1 (up to longitudinal--transverse interferences which are suppressed in most cases \cite{Denner:2021csi, Carrivale:2025mjy}), it can be interpreted as the probability for an event to have a longitudinally polarised boson.

As discussed above, the method was applied at \lo{} accuracy (possibly matched to PS), and extensions to higher-order corrections have not been addressed so far.
To obtain an approximate \rlltilde{} at \nlo{}, the ratio given in \cref{eq:rLL} has to be evaluated at this perturbative order. This requires evaluating, for a given phase-space point, two different matrix-element weights: one for a specific polarisation mode of the intermediate EW bosons and one for unpolarised intermediate EW bosons.
One approach to achieve this is to reweight a priori unpolarised events to events with a specific polarisation state.
{The implementation of such reweighting technique, followed by its verification and the subsequent regression and examination of \rlltilde{}, is the purpose of the present work.}

Before proceeding with the description of our approach in the presence of higher-order corrections, we remark that, while similar in spirit, the proposed strategy differs from the matrix-element method~\cite{Kondo:1988yd,Kondo:1991dw,Dalitz:1991wa,Dalitz:1992np,Kondo:1993in,Gao:2010qx,Bolognesi:2012mm,Anderson:2013afp,Gritsan:2016hjl,Gritsan:2020pib}
and its \nlo{} extensions~\cite{Alwall:2010cq,Campbell:2012cz,Campbell:2012ct,Campbell:2013hz,Martini:2015fsa,Baumeister:2016maz,Kraus:2019qoq,Kraus:2019myc,Martini:2023ylv,Tartarin:2025gbt,Tartarin:2026uoh,haisch2026matrixelementmethodnlo}.
While the matrix-element method typically constructs likelihood ratios from matrix elements corresponding to different underlying theory hypotheses, the amplitude-assisted method \cite{Grossi:2023fqq} makes use of polarised and unpolarised matrix elements to supervise the training of a flexible machine-learning classifier or regressor towards an efficient extraction of polarisation fractions.

\subsection{Reweighting approach with higher-order QCD corrections}\label{subsec:rew}
While the quantity defined in \cref{eq:rLL} is unambiguous at \lo{} in perturbation theory, the inclusion of higher-order corrections makes its definition more involved, owing to both virtual and real-radiation contributing to the same cross section beyond the \lo{}. By definition, such contributions are evaluated with different kinematics and necessarily need to be combined with suitable subtraction counterterms (local and integrated) to achieve infrared safety in a fully differential way.

In order to overcome this bottleneck, we employ the  
\POWHEG{}~method, a well-known approach to match \nlo{} calculations with parton showers \cite{Nason:2004rx,Frixione:2007vw,Alioli:2010xd}.
According to the \POWHEG{}~method, the calculation of a generic observable relies on the generation of the hardest emission originating from Born-like kinematics. Such kinematics are generated according to a \nlo{}-accurate weight, customarily denoted by $\bar{B}$, which is nothing but the NLO cross section, differential in the Born kinematics. As such, the $\bar{B}$ function  embeds IR-regulated one-loop virtual contributions and integrated unresolved real radiation, and it corrects the Born contribution to obtain \nlo{} accuracy:
\beq \label{eq:Bbar}
\bar{B}_{\lambda_1\lambda_2}(\tilde{\Phi}_4) = B_{\lambda_1\lambda_2} (\tilde{\Phi}_4) + V_{\lambda_1\lambda_2}^{\rm (reg)}(\tilde{\Phi}_4) + \int \! \rd\Phi_{\rm rad}\,  R^{\rm (reg)}_{\lambda_1\lambda_2} (\tilde{\bar{\Phi}}_4, \Phi_{\rm rad}) \,,
\eeq
where $B, V^{\rm reg}, R^{\rm reg}$ stand for the Born, IR-subtracted virtual, and IR-subtracted real contributions, respectively. All terms
in \cref{eq:Bbar} are fully differential in the Born phase space ($\Phi_4$),
and are evaluated in the DPA for the $\lambda_1,\lambda_2$ polarisation modes associated with the first and second EW boson, respectively. The tilded notation understands the application of the DPA to the original full off-shell Born kinematics ($\Phi_4\rightarrow\tilde{\Phi}_4$), while for the real phase space ($\Phi_5$) we assume the factorisation into a radiation phase space and a Born one, according to the subsequent application of subtraction mappings (${{\Phi}_5}\rightarrow {\bar{\Phi}}_4\Phi_{\rm rad}$) and of the
DPA on-shell projection (${\bar{\Phi}}_4\rightarrow \tilde{{\bar{\Phi}}}_4$)
\footnote{
Additional details on the \POWHEGBOXRES~implementation of di-boson processes in the DPA can be found in \citeres{Pelliccioli:2023zpd,Haisch:2025jqr}.
While simplified in the case of QCD initial-state radiation, which we consider here, the interplay between the DPA and subtraction mappings can be technically non-trivial in the presence of final-state radiation, as shown in fixed-order calculations \cite{Denner:2021csi,Le:2022lrp,Denner:2023ehn,Denner:2024tlu}.
}.
We stress that the application of the DPA (or any other on-shell approximation) to all \nlo{} terms is crucial to properly model intermediate EW bosons with definite polarisation states in the presence of higher-order corrections \cite{Denner:2020bcz}.
The sum over initial state partons and the convolution with PDFs is implicit in \cref{eq:Bbar}.
The IR-regulated real contribution is integrated over the phase space
of the additional radiation $\Phi_{\rm rad}$.
The no-emission probability above a certain value of the ordering variable $t$ is parametrised by a (\POWHEG{}) Sudakov form-factor $\Delta(t)$, which resums the leading logarithms from soft and collinear radiation. It is defined as:
\beq \label{eq:sudakov}
\Delta_{\lambda_1\lambda_2} (t) = \exp \left [ - \int_{t^\prime > t}\! \rd\Phi_{\rm rad}^\prime \, \frac{R_{\lambda_1\lambda_2} (\tilde{\bar{\Phi}}_4, \Phi_{\rm rad}^\prime)}{B_{\lambda_1\lambda_2} (\tilde{\Phi}_4)}\right ] \,.
\eeq
If only initial-state radiation (ISR) is present, as in the present case, $t$ equals the transverse momentum of the radiated parton in the partonic centre-of-mass frame.
For a generic IR-safe observable $ \mc O$, the hardest-emission generation reads then
\beq \label{eq:matching}
\langle {\cal O}\rangle = \int \! \rd\Phi_4 \, \bar{B}_{\lambda_1\lambda_2}(\tilde{\Phi}_4) \left [{\cal O}(\tilde{\Phi}_4) \, \Delta_{\lambda_1\lambda_2} (t_0) +\!\! \int_{t>t_0}\!\! \rd \Phi_{\rm rad}\, {\cal O}(\tilde{\bar{\Phi}}_4, \Phi_{\rm rad}) \, \frac{R_{\lambda_1\lambda_2} (\tilde{\bar{\Phi}}_4, \Phi_{\rm rad})}{B_{\lambda_1\lambda_2} (\tilde{\Phi}_4)}\, \Delta_{\lambda_1\lambda_2} (t) \right ] \, ,
\eeq
where $t_0$ is a cutoff of the order of $\Lambda_{\rm QCD}$. Subsequent emissions are softer and are generated by PS algorithms.
It can be shown that an observable computed according to \cref{eq:matching} is \nlo{} accurate.
In the case at hand, \POWHEGBOX~is matched to the leading-logarithmic \texttt{simple shower} of {\sc Pythia}8.2 \cite{Sjostrand:2014zea}.
For our purposes, we will only consider one QCD-parton emission in the hard process, \ie obtaining \nlo{} accuracy in QCD. 

Following the idea of \citere{Grossi:2023fqq} summarised in \cref{eq:fraction,eq:rLL}, we aim at constructing a quantity that describes fully differentially, \ie event by event, the relative weight between a configuration with two intermediate bosons with given polarisations $\lambda_1,\lambda_2$ and an unpolarised one. With these notations, at \lo{}, such a quantity is obviously given by the ratio between the lowest order squared matrix elements,
\beq \label{eq:rLL_def_lo}
r^{(\lo)}_{\lambda_1\lambda_2}(\tilde{\Phi}_4)
= \frac{{B}_{\lambda_1\lambda_2}(\tilde{\Phi}_4)}{{B}_{\rm unp}(\tilde{\Phi}_4)}\,,\qquad \lambda_1,\lambda_2 =\rL,\,\rT\,.
\eeq
However, at \nlo{} and beyond, both Born-like contributions and real-radiation contributions are present.
The analogue of \cref{eq:rLL_def_lo} at \nlo{} accuracy has to account for all contributions. The pure-Born weights are hence replaced by the $\bar{B}$ weights introduced in \cref{eq:Bbar}:
\beq \label{eq:rLL_def}
r^{(\nlo)}_{\lambda_1\lambda_2} (\tilde{\Phi}_4) = \frac{\bar{B}_{\lambda_1\lambda_2}(\tilde{\Phi}_4)}{\bar{B}_{\rm unp}(\tilde{\Phi}_4)}\,,\qquad \lambda_1,\lambda_2 =\rL,\,\rT\,.
\eeq
We observe that the ratio in \cref{eq:rLL_def} also contains contributions due to unresolved real radiation for specific polarisation states $\lambda_1,\lambda_2$, relatively to the unpolarised case. However, the \nlo{} description of a generic process also includes configurations where the real radiation is resolved (and potentially hard).
In the \POWHEGBOX~approach, this is precisely achieved by the second term inside the squared brackets of \cref{eq:matching}. In this sense, the ratio in \cref{eq:rLL_def} does not account for the polarisation configurations in hard real emissions exactly and in a fully differential manner.

The advantage of the $r_{\lambda_1\lambda_2}$ weights resides in the possibility
to reweight on-the-fly unweighted events generated with the \POWHEGBOX~approach for unpolarised bosons in order to obtain events with given polarised bosons. Starting from the unpolarised counterpart of \cref{eq:matching},
\beq \label{eq:matchingUNP}
\langle {\cal O}\rangle = \int \! \rd\Phi_4 \, \bar{B}_{\rm unp}(\tilde{\Phi}_4) \left [{\cal O}(\tilde{\Phi}_4) \, \Delta_{\rm unp} (t_0) + \int_{t>t_0}\! \rd \Phi_{\rm rad}\, {\cal O}(\tilde{\bar{\Phi}}_4, \Phi_{\rm rad}) \, \frac{R_{\rm unp} (\tilde{\bar{\Phi}}_4, \Phi_{\rm rad})}{B_{\rm unp} (\tilde{\Phi}_4)}\, \Delta_{\rm unp} (t) \right ] \,,
\eeq
the reweighting procedure proceeds, according to \cref{eq:rLL_def}, as follows
\beq\label{eq:replacement}
\bar{B}_{\rm unp}(\tilde{\Phi}_4)
\,\longrightarrow\,
r^{(\nlo)}_{\lambda_1\lambda_2}(\tilde{\Phi}_4) \,  \bar{B}_{\rm unp}(\tilde{\Phi}_4)
\equiv
\bar{B}_{\lambda_1\lambda_2}(\tilde{\Phi}_4)
\,.
\eeq
Such a reweighting procedure leads to an approximate version of \cref{eq:matching}, since the content of the square bracket of \cref{eq:matching} is computed using unpolarised matrix elements, rather than polarised ones. 
The goodness of this reweighting procedure can be assessed by comparing the results obtained in this approximation with the exact one of \cref{eq:matching}. We stress that while this reweighting approach is a standard for what concerns QCD-scale variations in the \POWHEGBOX framework, it is not guaranteed to be accurate in the case of helicity configurations. This is discussed in the next subsections.
We further stress that \rll{} is not guaranteed to be within zero and one for beyond-\lo{} events, due to the possibility of the weight $\bar{B}$ being negative.
The latter can happen, for example, in phase-space regions where large and negative virtual corrections overwhelm the positive sum of the Born and subtracted-real terms in \cref{eq:Bbar}.

The reweighting procedure described in this section has been implemented in the \POWHEGBOXRES~framework \cite{Alioli:2010xd,Jezo:2015aia} and is publicly available in the {\sc VV\!\_pol} package \cite{Pelliccioli:2023zpd,Haisch:2025jqr}.
The implementation also required some modifications of the source code in the matrix-element provider \recolatwo{} \cite{Denner:2017wsf}, which have been integrated in a recent public release
\footnote{
\recolatwo{} release 2.3.0, available at \url{https://gitlab.com/recola/recola2/-/releases/2.3.0}.
}.

\subsection{Numerical setup}
We consider the following di-boson process
\beq\label{eq:proc}
\Pp\Pp \to \PZ\PZ \to \Pe^+\Pe^- \hspace{0.5mm}\mu^+ \mu^-\,,
\eeq
at the LHC with a collision energy of $\sqrt{s}= 13 \, {\rm TeV}$.
The starting point of our approach is the simulation of unweighted events for the unpolarised processes, followed by the construction of 
a suitable reweighting approach that enables us to describe accurately the same processes but with intermediate longitudinally polarised bosons. Sticking to the general strategy of \citere{Grossi:2023fqq}, the used weights rely on the matrix element used to generate the di-boson processes with unpolarised and the polarised gauge bosons, respectively.

The simulations have been obtained for intermediate EW bosons
treated in the double-pole approximation \cite{Stuart:1991cc,Stuart:1991xk,Aeppli:1993cb,Aeppli:1993rs,Denner:2005fg,Denner:2019vbn}
with the {\sc VV\!\_pol} package \cite{Pelliccioli:2023zpd,Haisch:2025jqr}
in the \POWHEGBOXRES~framework \cite{Nason:2004rx,Frixione:2007vw,Alioli:2010xd,Jezo:2015aia}.
As SM input parameters, the on-shell masses and widths of EW bosons are taken from the latest PDG review~\cite{ParticleDataGroup:2024cfk} and are then converted to pole masses~\cite{Bardin:1988xt}, leading to
\begin{alignat}{2}
                \MZ &=  91.1535\GeV,      & \quad \quad \quad \GZ &= 2.4943\GeV,  \nonumber \\
                \MW &=  80.3499\GeV,       & \GW &= 2.0843\GeV .
\end{alignat}
We use the $G_\mu$ scheme, \ie the EW coupling $\alpha$ is computed from the real pole masses of weak bosons and the Fermi constant, which reads
\begin{align}
 \GF = 1.16638 \cdot 10^{-5}\, {\rm GeV}^{-2} .   
\end{align}
All leptons are treated as massless and the five-flavour scheme is employed. Quark mixing between generations is neglected. The proton PDFs and the QCD coupling constant $\alphas$ are evaluated through the \noun{Lhapdf} interface~\cite{Buckley:2014ana}, with the \sloppy {\tt NNPDF31\_nnlo\_as\_0118\_luxqed} set~\cite{NNPDF:2017mvq}. The factorisation and renormalisation scales are set to the arithmetic mean of the pole masses of the two bosons as
\begin{align}
 \mu_{\rm F} = \mu_{\rm R} = \MZ \,.
\end{align}
Throughout this study, parton showering is performed using \noun{Pythia~8.2}~\cite{Sjostrand:2014zea} and retaining QCD-shower effects only (the corresponding predictions are dubbed \nlops{}). The impact of leading-logarithmic QED-shower effects is rather polarisation-independent \cite{Pelliccioli:2023zpd,Carrivale:2025mjy} and therefore is not expected to change our conclusions.
The fixed-order (\nlo{}) EW corrections impact the polarisation extraction more significantly, although typically in suppressed regions of the phase-space \cite{Denner:2021csi,Carrivale:2025mjy}. A rigorous study of these effects goes beyond the purposes of this work and is therefore left for future investigations.

As for kinematic selections, we consider the fiducial setup corresponding to a recent ATLAS polarisation analysis 
\cite{ATLAS:2023zrv},
\begin{eqnarray}\label{eq:fidZZ}
&  81\GeV < M_{\ell^+\ell^-} <101\GeV\,, \qquad  M_{\rm 4\ell}>180\GeV\,,\qquad  {\rm \Delta} R_{\ell\ell'}>0.05\,,&\nnb\\
&  \pt{\Pe^\pm(\mu^\pm)}>7 (5)\GeV\,,\quad |y_{\Pe^\pm(\mu^\pm)}|<2.47(2.7)\,,\qquad \pt{\ell_{1(2)}}>20\GeV\,.&
\end{eqnarray}

\subsection{Impact of QCD radiation}
The inclusive production of EW-boson pairs at the LHC is known \cite{Kallweit:2019zez} to receive very large
QCD corrections, most of which come from hard real radiation.
As a consequence, it is unavoidable to include at least 
one QCD emission on top of the \lo{}, either by \lo{} multi-jet merging or by computing \nlo{} QCD corrections. This is especially crucial in the modelling of polarisation configurations, because the presence of additional QCD partons in the final state carrying spin quantum numbers makes room for new overall helicity configurations that may be suppressed at \lo{}. 
The impact of QCD radiation can be easily assessed by looking at the 
differential distributions in the transverse momentum of the colour singlet, \ie the four-lepton system in the processes under consideration.
\begin{figure}
\centering
\includegraphics[width=.49\textwidth,page=7]{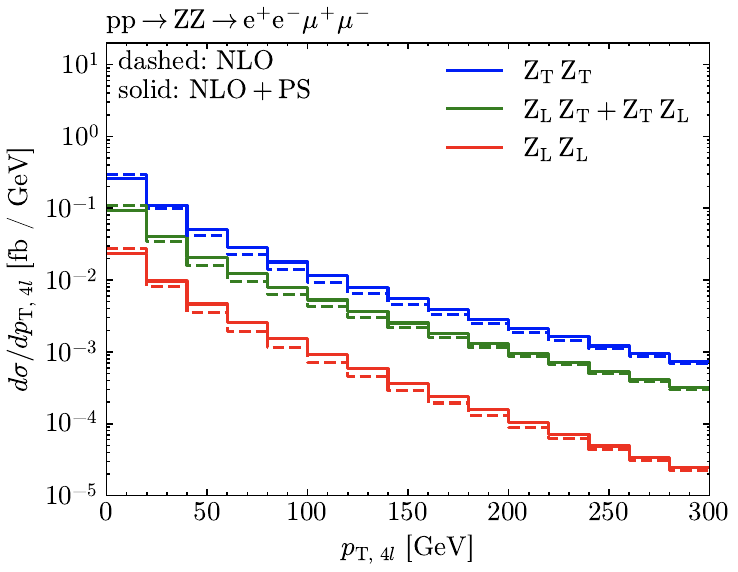}
\caption{Differential distributions in the transverse momentum of the four-lepton system for doubly polarised $\PZ\PZ$ production at the LHC at $13\TeV$ in the ATLAS fiducial setups of \citere{ATLAS:2023zrv}.
The results are \nlo{} QCD accurate and are shown at fixed order (dashed curves) and matched to QCD parton shower (solid curves). 
}\label{fig:ZZ_pts}
\end{figure}
These results are shown in \cref{fig:ZZ_pts} for $\PZ\PZ$ production at the LHC in fiducial setups that mimic recent polarisation analyses by ATLAS \cite{ATLAS:2023zrv}. The results in a more inclusive setup ($81\GeV<M_{\ell^+\ell^-}<101\GeV$ cuts only) are largely equivalent, up to minor effects due to the omitted transverse-momentum and rapidity cuts on individual leptons.
At \nlo{} QCD, the differential distribution of the transverse momentum of the colour singlet ($\pt{4\ell}$) only receives contributions from real radiation, making it, effectively, \lo{} accurate. The difference between \nlo{} and \nlops{} results shown in \cref{fig:ZZ_pts} is easily understood by recalling that the emission of multiple soft and collinear radiation dampens the results at small $\pt{4\ell}$, where a Sudakov peak is present, thereby making the tails slightly harder.

As already shown in the literature \cite{Pelliccioli:2023zpd,Haisch:2025jqr}, pairs of longitudinal bosons are produced in softer configurations, \ie at lower $\pt{4\ell}$ compared to purely transverse and mixed polarisation states, with hard-real contributions being more suppressed than in other states. This can be appreciated from the faster fall-off of the $\rL\rL$ curves compared to the other ones, both at fixed order (\nlo{} QCD) and after matching to PS.
On the one hand, having the smallest effect of hard QCD radiation for the $\rL\rL$ polarisation state, we expect that the error we made by keeping the unpolarised parts of \cref{eq:matchingUNP} when performing the replacement in \cref{eq:replacement} should have a reasonably small impact.
On the other hand, the $\rT\rT$ state contributes 65\% to the unpolarised integrated cross section and drives the unpolarised modelling of the di-boson process both in the Born-like and in the hard-radiation contributions, leading to a small mismatch in the Sudakov exponential.
This suggests that the reweighting approach according to $r_{\rT\rT}$ within the \POWHEGBOX{} method should also perform well too in reproducing the results of the direct simulation of the $\rT\rT$ signal. Indeed, these expectations are confirmed by the results discussed in \cref{subsec:tests}.

Finally, we note that the inclusion of NNLO QCD effects partially enhances the $\rL\rL$ tails in $\pt{4\ell}$ \cite{Carrivale:2025mjy,Pelliccioli:2025com}. 
However, the overall behaviour of the $\rL\rL$ distribution remains the same as at \nlo{} QCD, with a stronger suppression compared to the other polarisation states. 
While extending our analysis to NNLO QCD accuracy is desirable once the generation of polarised-boson signals will be possible in the {\sc MiNNLOPS} framework \cite{Monni:2019whf}, we are confident that the conclusions of the present work should not be changed substantially by the inclusion of NNLO QCD corrections.

\subsection{Closure tests}\label{subsec:tests}
Before using the weights in \cref{eq:rLL_def} as labels for LHC events to train a NN model, we need to 
assess the goodness of the reweighting technique introduced in \cref{subsec:rew} in the \POWHEG{} framework in reproducing results that are directly simulated for polarised intermediate bosons. This has to be done both for integrated and 
differential cross sections. The ATLAS setup of \citere{ATLAS:2023zrv} is considered throughout this subsection.

The results for the fiducial cross section are shown in \cref{tab:int-rew} for various different accuracies. At \lo{}, the $\bar{B}$ of \cref{eq:Bbar} reduces to $B$. For \lows{} resummation effects from the Sudakov form factor in \cref{eq:sudakov} are considered as well. \nlo{}$_{\rm QCD}$ (LHE), denotes the fiducial cross section, which is coming from events where only the hardest radiation generated by \POWHEG{} (the so-called "LHE" level) is kept and, finally, for \nlops$_{\rm QCD}$ the events are additionally matched to the {\sc Pythia} 8.2 QCD shower.
The reweighting performs very well for all polarisation states at \lo{}, while the inclusion of \nlo{} effects and the matching to PS makes the reweighting for mixed states underestimate the cross section by $7\%$.
At \nlops{}, the reweighting for $\rL\rL$ and $\rT\rT$ works pretty well, reproducing the simulated integrated results by $-0.9\%$ and $+0.3\%$, respectively.
{\renewcommand{\arraystretch}{1.5}
\begin{table}
\begin{center}
\begin{tabular}{l|cc|cc|cc}
& \multicolumn{2}{c|}{\phantom{$\rT$}$\rL\rL$\phantom{$\rT$}} &  \multicolumn{2}{c|}{$\rL\rT+\rT\rL$} & \multicolumn{2}{c}{\phantom{$\rT$}$\rT\rT$\phantom{$\rT$}}  \\
& sim. & rew. & sim. & rew. & sim. & rew. \\
\hline
\hline
\lo{}                    & $ 0.65751(8)$ & $0.6573(4) $ & $ 2.6707(2) $ & $2.6695(8) $ & $ 7.7887(8) $ & $7.790(2)  $ \\
\lows{}                  & $ 0.6634(2) $ & $0.6657(4) $ & $ 2.6949(6) $ & $2.6968(8) $ & $ 7.833(3)  $ & $7.870(2)  $ \\
\nlo{}$_{\rm QCD}$ (LHE) & $ 0.8902(1) $ & $0.8841(5) $ & $ 3.8554(4) $ & $3.608(1)  $ & $ 10.211(1) $ & $10.240(3) $ \\
\nlops$_{\rm QCD}$       & $ 0.8918(3) $ & $0.8841(5) $ & $ 3.8660(9) $ & $3.608(1)  $ & $ 10.215(3) $ & $10.241(3) $ \\
\end{tabular}\qquad
\end{center}
\vspace{0mm}
\caption{
Fiducial cross sections (in fb) for doubly polarised $\PZ\PZ$ production at the LHC@13TeV in the ATLAS fiducial setups of \citere{ATLAS:2023zrv}. The simulated signals are compared to the results obtained with the reweighting approach described in \cref{subsec:rew}. Numerical integration uncertainties are shown in parentheses.
\label{tab:int-rew}}
\end{table}
}

\begin{figure}[htb!]
\centering
\includegraphics[height=.41\textheight,page=1]{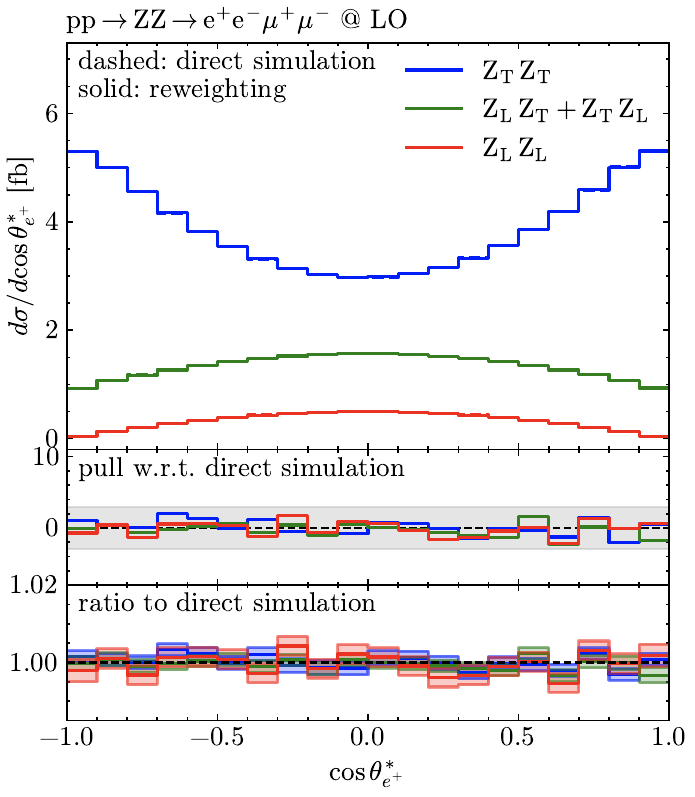} \hfill
\includegraphics[height=.41\textheight,page=1]{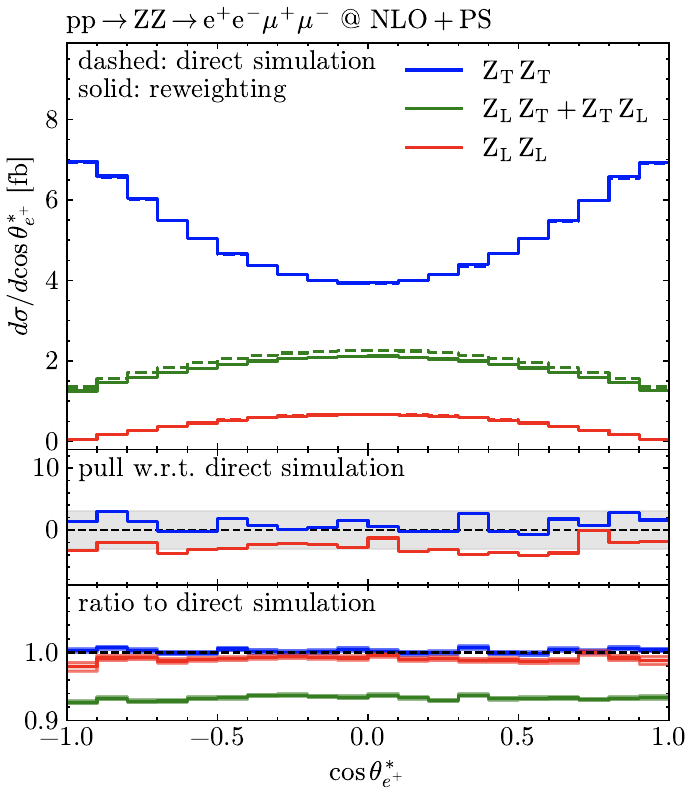}
\caption{Differential distributions with respect to the positron polar decay angle for doubly polarised $\PZ\PZ$ production at the LHC@13TeV in the ATLAS fiducial setups of \citere{ATLAS:2023zrv}.
The simulated signals (solid curves) are compared to the results obtained with the reweighting approach described in \cref{subsec:rew} (dashed curves). 
The results are shown at fixed-order \lo{} accuracy (left panel) and at \nlo{} QCD accuracy matched to QCD PS (right panel). 
Same colour key as in \cref{fig:ZZ_pts}. Upper panels: absolute distributions. Middle panels: 3$\sigma$ pulls (reweighted vs simulated, from numerical-integration uncertainties). Lower panels: ratios of reweighted distributions to direct simulations. The bands represent the MC error of the predictions.
}
\label{fig:ZZ_rew_LO_costheta}
\end{figure}

\begin{figure}[htb!]
\centering
\includegraphics[height=.41\textheight,page=2]{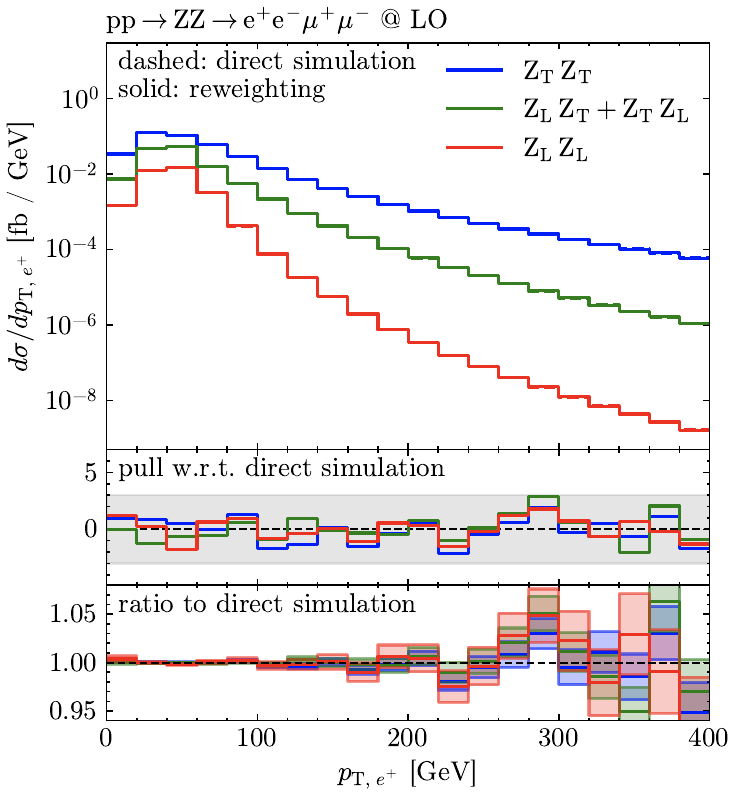} \hfill
\includegraphics[height=.4\textheight,page=2]{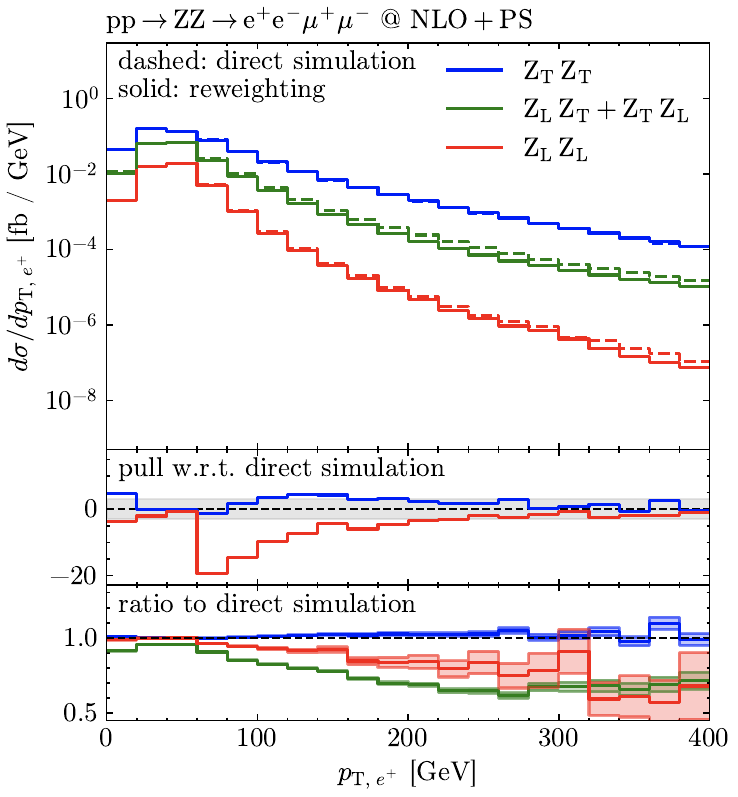}
\caption{
Same as \cref{fig:ZZ_rew_LO_costheta} but for the positron transverse momentum. 
}
\label{fig:ZZ_rew_LO_ptep}
\end{figure}

In \cref{fig:ZZ_rew_LO_costheta,fig:ZZ_rew_LO_ptep}, we show the differential results for two different LHC observables. We show both the different contributions from the $\rL\rL$, $\rT\rT$, and mixed polarisation signals, as well as predictions at LO and NLO+PS accuracy.
In the middle panels we show the pull, \ie the difference of the reweighting results with respect to the directly simulated predictions over their combined MC uncertainty. The lowest panel displays the ratio of the two predictions, with MC uncertainties indicated by shaded bands.


The polar decay angle of the positron in the corresponding $\PZ$-boson rest frame ($\theta^*_{\Pe^+}$) represents the most decay-like observable which can be reconstructed in $\PZ\PZ$ fully leptonic events, and is sensitive to the single-boson polarisation state.
The positron transverse momentum ($\pt{\Pe^+}$) is an energy-dependent observable, whose tail encodes the regime where the dynamics of longitudinal modes of EW bosons resemble those of the corresponding Goldstone bosons \cite{Cornwall:1974km,Vayonakis:1976vz}. However, this suppressed region is hard to probe at the LHC owing to limited statistics \cite{ATLAS:2023zrv}.

At \lo{} accuracy, the reweighting approach is, as expected, exact and differs from the direct simulation only at the statistical level.
The same conclusion holds true at \nlops{} for the LL and TT polarised predictions for the distribution in $\cos \theta^*_{\Pe^+}$.
For the LT+TL predictions, while the shape is well described, the normalisation differs by about $7\%$, a pattern similar to what we observe for the total cross section.
For the transverse momentum of the positron, instead, the reweighting approach is good at the per-cent level over the full range at LO.
For LL signals, the reweighted prediction agrees within $5\%$ until $150 \GeV$, where most events lie.
Above this energy, the direct simulation and the reweighted prediction depart from one another, with a low impact on the total cross section, due to their kinematic suppression.
It is worth noting that at \nlops{} accuracy and around $60\GeV$ the pull diverges significantly while it smoothly reduces towards higher energies.
This is caused by the very good agreement between the central values in the first three bins, together with very small MC uncertainties in both predictions. When the reweighting prediction in the fourth bin at $60\GeV$ starts to disagree slightly with the simulated prediction, the MC uncertainty is still rather small, resulting in a large pull in the middle panel. 
Going to higher energy, the difference is increasing, but the statistical error is increasing even faster, explaining the behaviour of the pull.
Finally, the LT+TL predictions only agree in the very bulk of the distribution \ie around $50\GeV$.

\section{Regression approaches}\label{sec:ml}
While the current paradigm of ATLAS and CMS polarisation-targeted analyses is a template fit with polarised sub-signals simulated independently with standard MC techniques
\cite{Aaboud:2019gxl,CMS:2020etf,CMS:2021icx,ATLAS:2022oge,ATLAS:2023zrv,ATLAS:2024qbd,ATLAS:2025wuw}, it is of interest to investigate novel approaches which could make the extraction of polarisation fractions and polarised cross sections (both integrated and differential) more efficient. 
In this work, we exploit a regression approach for the ratio \rll{}, which is most directly connected to the pseudo-observables of interest, namely the joint longitudinal-polarisation fractions.
As mentioned before, the target label \rll{} could also be used to perform a polarisation tagging of di-boson events.

In the present work, we examine multiple regression strategies to determine \rll{}:
\begin{itemize}
\item an adapted version of the feed-forward NN employed in \citere{Grossi:2023fqq} (dubbed \ffnn{}),
\item an autoencoder-like NN (\autoencoder{}),
\item a physics-informed NN (\pn{}), and
\item a random-forest regressor (\rfr{}).
\end{itemize}
The most agnostic choice for the input features is the complete set of four momenta of the charged leptons, written in the laboratory reference frame in Cartesian coordinates,
\begin{equation}
    \label{eq:input17}
    E_{\Pe^+},\, \Vec{p}_{\Pe^+},\quad
    E_{\Pe^-},\, \Vec{p}_{\Pe^-},\quad
    E_{\mu^+},\, \Vec{p}_{\mu^+},\quad
    E_{\mu^+},\, \Vec{p}_{\mu^+}
    \, ,
\end{equation}
which is combined with the corresponding target \rll{}.
We note that this input set, built out of 16 features, is actually redundant.
For example, charged leptons are massless, which could reduce the number of features by four (one per final-state particle). 
We therefore investigate another commonly used parametrisation (see \emph{e.g.} Refs.~\cite{Komiske:2018cqr,Qu:2019gqs}) of the two reconstructed Z bosons as input for the NNs, consisting of the $\PZ$-boson mass, transverse momentum, rapidity, and azimuthal angle in the laboratory frame.
To uniquely parametrise the lepton kinematics, the polar and azimuthal decay angles of the charged leptons in the respective $\PZ$-boson rest frame are needed:
    \beq\label{eq:input_prod_dec}
    M_{\PZ_1},\, \pt{\PZ_1},\, y_{\PZ_1},\, \phi_{\PZ_1},\quad 
    M_{\PZ_2},\, \pt{\PZ_2},\, y_{\PZ_2},\, \phi_{\PZ_2},\quad 
    \cos\theta^*_{\Pe^+},\, \phi^*_{\Pe^+},\quad 
    \cos\theta^*_{\mu^+},\, \phi^*_{\mu^+}.
    \eeq
This set of features (12) makes the separation between production-level and decay-level quantities explicit and relies on the massless nature of the leptons.
It is worth mentioning that it includes the, for polarisation states, very sensitive, polar decay angles $\cos\theta^*_{\Pe^+ / \, \mu^+}$.
Note that for the \pn{} and \rfr{} models, we have also considered alternative choices of input features, \emph{e.g.}, exploiting a combination of observables written in different Lorentz reference frames.

\subsection{Feed-forward NN}\label{subsec:ffnn}
In this section, we consider a fully connected, feed-forward NN (dubbed \ffnn{}) similar to the one utilised in \citere{Grossi:2023fqq}. The architecture of the network employed there was chosen as a trade-off between having a high level of non-linear expressivity and having a well-trainable network.

We start from the architecture of \citere{Grossi:2023fqq} applied to 16 input features. The input data is standardised first, ensuring an arithmetic mean of one and a vanishing standard deviation for each input feature. The standardisation is followed by a linear layer with dimensionality $\mathbb{R}^{16 \times 1000}$ and a rectified linear unit (\relu{}) \cite{glorot2011deep}. 
The main hidden block is built from 4 pairs of a linear layer with dimension $\mathbb{R}^{1000 \times 1000}$ and another \relu{}. Finally, a layer with dimension $\mathbb{R}^{1000 \times 1}$ reduces the network output to a single estimator for \rll{}.
Each linear layer is fully connected and includes a bias with appropriate dimensions, resulting in a network with approximately 4M parameters.

Owing to the larger number of kinematic features (16) for the $\PZ\PZ$ process compared to those for $\PZ\Pj$ (12) considered in \citere{Grossi:2023fqq}, as well as to the higher accuracy of the events used to predict \rll{} (up to \nlops{} compared to \lops{}), a lower width-to-depth ratio, which improves the network expressivity, was needed.

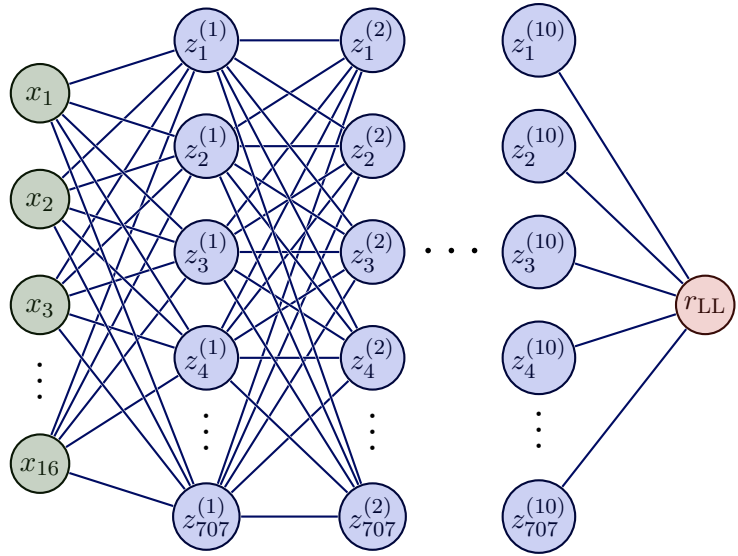
\begin{figure}[tb]
    \centering
    \colorlet{myred}{red!80!black}
\colorlet{myblue}{blue!80!black}
\colorlet{mygreen}{green!60!black}
\colorlet{mydarkred}{red!30!black}
\colorlet{mydarkblue}{blue!40!black}
\colorlet{mydarkgreen}{green!30!black}

\tikzset{
  >=latex, 
  node/.style={thick,circle,draw=myblue,minimum size=22,inner sep=0.5,outer sep=0.6},
  node in/.style={node,green!20!black,draw=mygreen!30!black,fill=mygreen!25},
  node hidden/.style={node,blue!20!black,draw=myblue!30!black,fill=myblue!20},
  node out/.style={node,red!20!black,draw=myred!30!black,fill=myred!20},
  connect/.style={thick,mydarkblue}, 
  connect arrow/.style={-{Latex[length=4,width=3.5]},thick,mydarkblue,shorten <=0.5,shorten >=1},
  node 1/.style={node in}, 
  node 2/.style={node hidden},
  node 3/.style={node out}
}
\def\nstyle{int(\lay<\Nnodlen?min(2,\lay):3)} 

\begin{tikzpicture}[x=2.2cm,y=1.4cm]
  \message{Same width feed forward neural network}
  \readlist\Nnod{4,5,5,5,1} 
  \readlist\Nstr{16,707,707,707,}
  \readlist\Cstr{\strut x,z^{(\prev)},z^{(\prev)},z^{(10)},r_{\mathrm{LL}}}
  \def\yshift{0.5}

  \message{^^J  Layer}
  \foreachitem \N \in \Nnod{
    \def\lay{\Ncnt}
    \pgfmathsetmacro\prev{int(\Ncnt-1)}
    \message{\lay,}

    \foreach \i [evaluate={\c=int(\i==\N); \y=\N/2-\i-\c*\yshift;
                 \index=(\i<\N?int(\i):"\Nstr[\lay]");
                 \x=\lay; \n=\nstyle;}] in {1,...,\N}{ 

        \ifnum\lay<\Nnodlen
            \node[node \n] (N\lay-\i) at (\x,\y) {$\Cstr[\lay]_{\index}$};
        \else
            \node[node \n] (N\lay-\i) at (\x,\y) {$\Cstr[\lay]$};
        \fi

        \ifnum\lay>1
            \ifnum\lay=4
    
            \else
              \foreach \j in {1,...,\Nnod[\prev]}{
                \draw[connect,white,line width=1.2] (N\prev-\j) -- (N\lay-\i);
                \draw[connect] (N\prev-\j) -- (N\lay-\i);
              }
            \fi
        \fi
    }

    \ifnum\lay<\Nnodlen
        \path (N\lay-\N) --++ (0,1+\yshift) node[midway,scale=1.5] {$\vdots$};
    \fi
    
    \ifnum\lay=3
      \node at ($(N3-3)+(0.5,0)$) [scale=1.8] {$\cdots$};
    \fi
  }

\end{tikzpicture}
    \caption{Architecture of our \ffnn{} model, namely a NN with 10 hidden layers and a constant width of 707. Picture obtained with {\tt tikz} \cite{nnscheme}.}
    \label{fig:ffnn_architecture}
\end{figure}

Specifically, we find that using a \ffnn{} with 4 additional pairs of a linear layer and a \relu{} in the hidden block, while keeping the total number of \ffnn{} parameters fixed, improves the learning capability. The network architecture with a correspondingly reduced width of 707 is shown in \cref{fig:ffnn_architecture}.
Nonetheless, such a network, with a lower width-to-depth ratio, can be harder to train \cite{glorot2011deep} compared to the original one \cite{Grossi:2023fqq}.
For example, this choice of network can result in a constant prediction $\tilde{r}_{\rL\rL}$ for the target label, 
\begin{equation}
    \tilde{r}_{\rL\rL}\approx \frac{\sigma_{\rL \rL}^{\rm fid}}{\sigma_{\mathrm{unp}}^{\rm fid}}
    \, ,
\end{equation}
corresponding to the mean value of the dataset.
This originates from the fact that learning the mean value for \rll{} leads, on average, to a relatively small loss value at the cost of failing to capture the full dynamic of \rll{} over the full phase space.
To avoid this issue and have more stable gradients at the training level, a $\log$-transformation,
\begin{equation}
    \label{eq:logtrafo}
    \rll \longrightarrow \log (1 + \rll)
    \, ,
\end{equation}
is applied to the target during the training, and then inverted when the training is complete.
This reduces the chances for the network to collapse into a constant prediction of \rll{}, since the scale of large values and thus the scale of large gradients itself is reduced.
The transformation in \cref{eq:logtrafo} ensures a well-defined $\log$-transformation even when small negative values of \rll{} occur, as described at the end of \cref{subsec:rew}.

At variance with the architecture of \citere{Grossi:2023fqq}, we
have replaced the \emph{a priori} standardisation of the input data with a batch~normalisation~\cite{ioffe2015batch}, standardising each batch in the training separately, while allowing the network to unlearn this normalisation with 32 additional, trainable parameters.
Starting with a complete set of 4.7M di-boson events, 
the training is performed using \noun{PyTorch}~\cite{ansel2024pytorch2} on $60 \%$ of the events, while the remaining events are used for validation (20\%) and testing (20\%).
During training, we exploit the \noun{AdamW} optimiser~\cite{loshchilov2018decoupled} with a small weight decay of $10^{-4}$, allowing to reduce over-fitting.
A second measure to prohibit over-fitting is taken by introducing a patience parameter (fixed empirically to a value of 35). After each epoch, the loss is compared between the training dataset and the validation dataset. If the validation loss is bigger than the training loss, a patience counter is increased and the training is stopped when the maximum patience is reached.
An illustration of this behaviour can be observed in \cref{fig:training_history} of \cref{app:hyperparameters}.
Finally, by manually varying the hyperparameters, we observe that the batch size and the learning rate are the hyperparameters with the greatest impact.
The hyperparameters have been optimised using the \noun{Optuna} framework~\cite{akiba2019optunanextgenerationhyperparameteroptimization}.
The optimal values for \lo{}, \lows{} and \nlops{} are summarised in \cref{tab:best_hyperparameters} of \cref{app:hyperparameters}.
Although the hyperparameter optimisation gives slightly different optimal values for the various accuracies, we have checked that using the \lo{} settings for the \lows{} and \nlops{} training does not lead to physically relevant differences in the predictions.

\subsection{Autoencoder-like NN}\label{subsec:aenn}
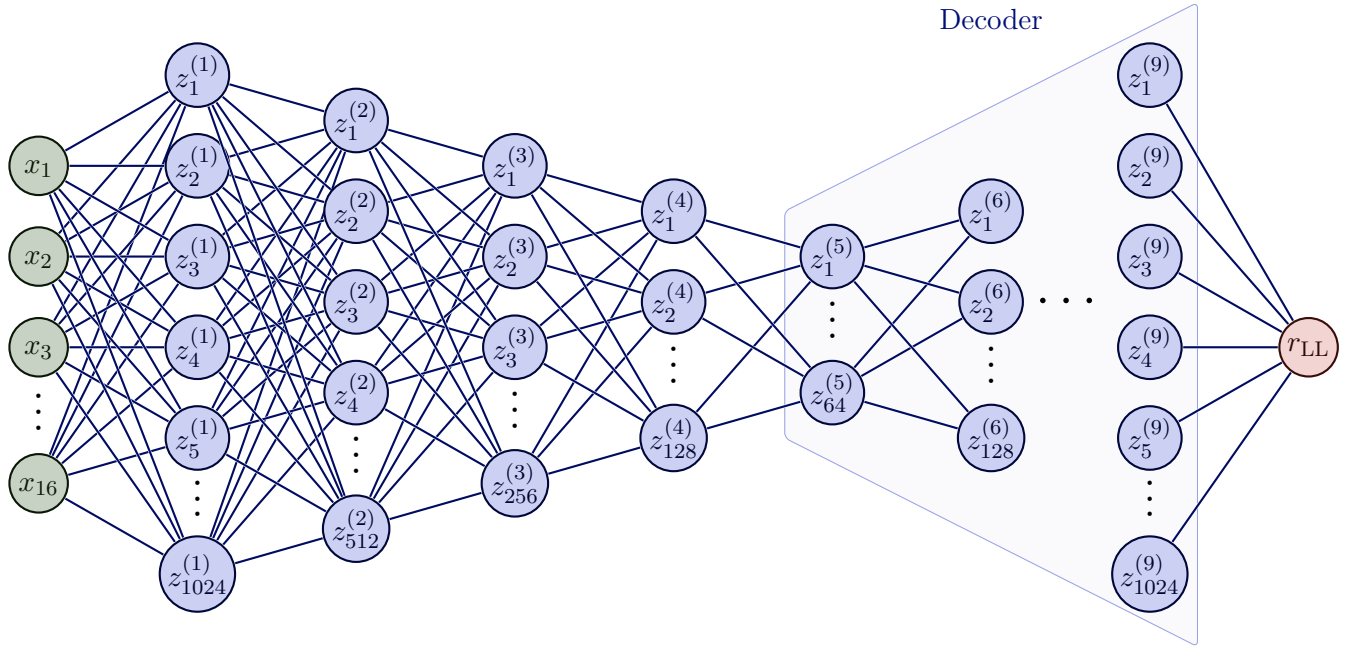
\begin{figure}[tb]
    \centering
    \colorlet{myred}{red!80!black}
\colorlet{myblue}{blue!80!black}
\colorlet{mygreen}{green!60!black}
\colorlet{mydarkred}{red!30!black}
\colorlet{mydarkblue}{blue!40!black}
\colorlet{mydarkgreen}{green!30!black}

\tikzset{
  >=latex, 
  node/.style={thick,circle,draw=myblue,minimum size=22,inner sep=0.5,outer sep=0.6},
  node in/.style={node,green!20!black,draw=mygreen!30!black,fill=mygreen!25},
  node hidden/.style={node,blue!20!black,draw=myblue!30!black,fill=myblue!20},
  node out/.style={node,red!20!black,draw=myred!30!black,fill=myred!20},
  connect/.style={thick,mydarkblue}, 
  connect arrow/.style={-{Latex[length=4,width=3.5]},thick,mydarkblue,shorten <=0.5,shorten >=1},
  node 1/.style={node in}, 
  node 2/.style={node hidden},
  node 3/.style={node out}
}
\def\nstyle{int(\lay<\Nnodlen?min(2,\lay):3)} 

\begin{tikzpicture}[x=2.1cm,y=1.2cm]
    \message{^^JNeural network without arrows}
    \readlist\Nnod{4,6,5,4,3,2,3,6,1} 
    \readlist\Nstr{16,1024,512,256,128,64,128,1024,}
    \readlist\Cstr{\strut x,z^{(\prev)},z^{(\prev)},z^{(\prev)},z^{(\prev)},z^{(\prev)},z^{(\prev)},z^{(9)},r_{\mathrm{LL}}}
    \def\yshift{0.5}

    \node[above,align=center,myblue!60!black] at (7,2.4) {Decoder};
    \draw[myblue!40,fill=myblue,fill opacity=0.02,rounded corners=2] (8.3,-4.3) --++ (0,7.1) --++ (-2.6,-2.3) --++ (0,-2.5) -- cycle;

    \message{^^J  Layer}
    \foreachitem \N \in \Nnod{ 
        \def\lay{\Ncnt} 
        \pgfmathsetmacro\prev{int(\Ncnt-1)} 
        \message{\lay,}

        \foreach \i [evaluate={\c=int(\i==\N); \y=\N/2-\i-\c*\yshift;
                     \index=(\i<\N?int(\i):"\Nstr[\lay]");
                     \x=\lay; \n=\nstyle;}] in {1,...,\N}{ 
            \ifnum\lay<\Nnodlen
                \node[node \n] (N\lay-\i) at (\x,\y) {$\Cstr[\lay]_{\index}$};
            \else
                \node[node \n] (N\lay-\i) at (\x,\y) {$\Cstr[\lay]$};
            \fi

            \ifnum\lay>1 
                \ifnum\lay=8
                \else
                    \foreach \j in {1,...,\Nnod[\prev]}{ 
                      \draw[connect,white,line width=1.2] (N\prev-\j) -- (N\lay-\i);
                      \draw[connect] (N\prev-\j) -- (N\lay-\i);
                    }
                \fi
            \fi 
        }

        \ifnum\lay<\Nnodlen
            \path (N\lay-\N) --++ (0, 1+\yshift) node[midway, scale=1.5] {$\vdots$};
        \fi

        \ifnum\lay=7
          \node at ($(N7-2)+(0.5,0)$) [scale=1.8] {$\cdots$};
        \fi
    }

\end{tikzpicture}
    \caption{Structure of the autoencoder-like NN (\autoencoder{}). The decoder part of the network on the right, not shown, corresponds to the mirroring of the encoder part on the left. 
    Picture obtained with {\tt tikz}~\cite{nnscheme}.}
    \label{fig:autoencoder_architecture}
\end{figure}

The core task of the NN is to convert the 16 input features into a single prediction for \rll{}.
It is therefore natural to consider, as a second architecture, a network with a built-in bottleneck structure, such as an autoencoder, which encourages the network to learn how to compress the data most effectively. Starting from 16 features, we increase the width of the network to 1024 nodes and then reduce it to 64 over 5 layers. Each layer is followed by a \relu{}. This encoder part of the network is then inverted by a so-called decoder, which iteratively increases the network's width back to 1024 nodes. A final linear layer reduces all parameters back to the unique output \rll{}. The architecture of such an autoencoder-like NN (dubbed \autoencoder{}) is illustrated in \cref{fig:autoencoder_architecture}.

The training setup is the same as in \cref{subsec:ffnn}, including the logarithmic transformation of the target, as shown in \cref{eq:logtrafo}. Hyperparameter fitting is performed in a fully independent manner compared to the \ffnn{} described previously.
The best values for the batch size and learning rate are provided in \cref{tab:best_hyperparameters} of \cref{app:hyperparameters} together with those of the \ffnn{} model.

\subsection{Physics-informed model} \label{subsec:pn}

Since we rely on simulations for which we can generate arbitrarily large amounts of data, the approach described in the previous section has focused on obtaining the best possible description of \rll{} regardless of the size of the network.
This is an effective approach when the amount of data is large enough such that a large network can learn the underlying physical representation of the data~\cite{Vigl:2026ppx}, and when the available computing power is sufficient to accommodate both the model size and the data size. This has been described as a scaling phenomenon in the field of Large Language Models~\cite{2020arXiv200108361K}, which recently has garnered a lot of interest also in particle physics~\cite{Bahl:2026jvt,Vigl:2026ppx}.

However, as the size of the dataset and the complexity of the problem increase, the network architecture might need to grow proportionally.
This growth can introduce a computational bottleneck, limiting the practical usage of this methodology as well as future extensions.
In order to pre-empt this issue, we complete our study of NN models by incorporating physical information directly into the design of the model with the goal of simplifying the architecture without a penalty on performance.

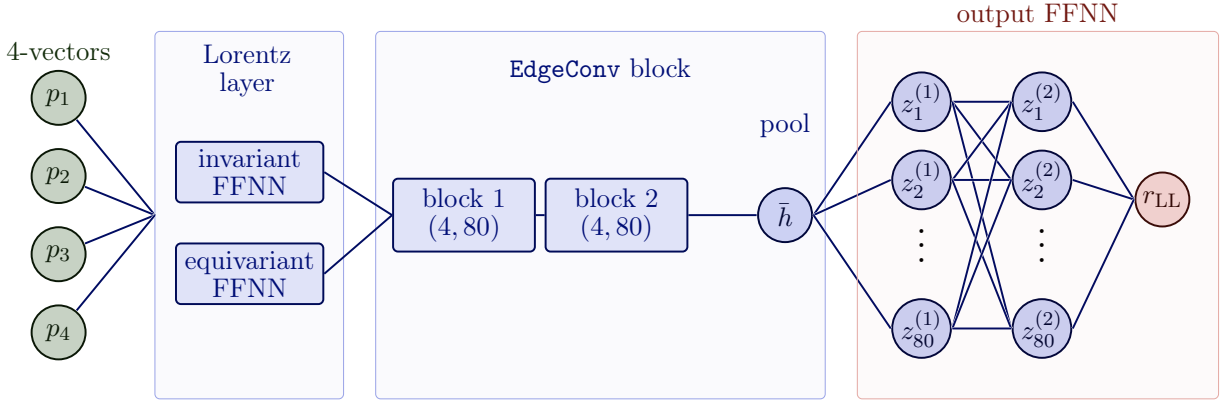
\begin{figure}[tb]
    \centering
    \resizebox{.95\textwidth}{!}{\colorlet{myred}{red!80!black}
\colorlet{myblue}{blue!80!black}
\colorlet{mygreen}{green!60!black}
\colorlet{mydarkred}{red!30!black}
\colorlet{mydarkblue}{blue!40!black}
\colorlet{mydarkgreen}{green!30!black}

\tikzset{
  >=latex, 
  node/.style={thick,circle,draw=myblue,minimum size=22,inner sep=0.5,outer sep=0.6},
  node in/.style={node,green!20!black,draw=mygreen!30!black,fill=mygreen!25},
  node hidden/.style={node,blue!20!black,draw=myblue!30!black,fill=myblue!20},
  node out/.style={node,red!20!black,draw=myred!30!black,fill=myred!20},
  connect/.style={thick,mydarkblue},
  connect arrow/.style={-{Latex[length=4,width=3.5]},thick,mydarkblue,shorten <=0.5,shorten >=1}
}

\begin{tikzpicture}[x=1.15cm,y=1.12cm]
  \message{ParticleNet architecture}
  \def\yshift{0.5}

  \foreach \i [evaluate={\y=2.5-\i;}] in {1,...,4}{
    \node[node in] (X-\i) at (1,\y) {$p_{\i}$};
  }

  \draw[myblue!40,fill=myblue,fill opacity=0.02,rounded corners=2]
    (2.20,-2.35) rectangle (4.55,2.35);
  \node[align=center,myblue!60!black] at (3.35,1.85)
    {Lorentz\\[-0.2em]layer};
  \node[
    draw=myblue!45!black,
    fill=myblue!12,
    thick,
    rounded corners=2,
    align=center,
    text width=1.9cm,
    inner sep=3pt,
    text=myblue!60!black
  ] (Inv) at (3.38,0.55) {invariant\\[-0.2em]\ffnn{}};
  \node[
    draw=myblue!45!black,
    fill=myblue!12,
    thick,
    rounded corners=2,
    align=center,
    text width=1.9cm,
    inner sep=3pt,
    text=myblue!60!black
  ] (Eq) at (3.38,-0.75) {equivariant\\[-0.2em]\ffnn{}};

  \draw[myblue!40,fill=myblue,fill opacity=0.02,rounded corners=2]
    (4.95,-2.35) rectangle (10.55,2.35);
  \node[align=center,myblue!60!black] at (7.75,1.85)
    {\texttt{EdgeConv} block};
  \node[
    draw=myblue!45!black,
    fill=myblue!12,
    thick,
    rounded corners=2,
    align=center,
    text width=1.75cm,
    inner sep=4pt,
    text=myblue!60!black
  ] (EC1) at (6.05,0) {block 1\\[-0.2em]$(4,80)$};
  \node[
    draw=myblue!45!black,
    fill=myblue!12,
    thick,
    rounded corners=2,
    align=center,
    text width=1.75cm,
    inner sep=4pt,
    text=myblue!60!black
  ] (EC2) at (7.95,0) {block 2\\[-0.2em]$(4,80)$};

  \node[node hidden] (Pool) at (10.05,0) {$\bar h$};
  \node[node hidden] (Z1-1) at (11.75,1.45) {$z^{(1)}_1$};
  \node[node hidden] (Z1-2) at (11.75,0.45) {$z^{(1)}_2$};
  \node at (11.75,-0.25) [scale=1.45] {$\vdots$};
  \node[node hidden] (Z1-80) at (11.75,-1.45) {$z^{(1)}_{80}$};
  \node[node hidden] (Z2-1) at (13.25,1.45) {$z^{(2)}_1$};
  \node[node hidden] (Z2-2) at (13.25,0.45) {$z^{(2)}_2$};
  \node at (13.25,-0.25) [scale=1.45] {$\vdots$};
  \node[node hidden] (Z2-80) at (13.25,-1.45) {$z^{(2)}_{80}$};
  \node[node out] (Out) at (14.75,0.20) {$r_{\mathrm{LL}}$};

  \foreach \i in {1,...,4}{
    \draw[connect,white,line width=1.2] (X-\i) -- (2.20,0);
    \draw[connect] (X-\i) -- (2.20,0);
  }
  \draw[connect] (Inv.east) -- (EC1.west);
  \draw[connect] (Eq.east) -- (EC1.west);
  \draw[connect] (EC1.east) -- (EC2.west);
  \draw[connect] (EC2.east) -- (Pool.west);
  \foreach \i in {1,2,80}{
    \draw[connect,white,line width=1.2] (Pool.east) -- (Z1-\i.west);
    \draw[connect] (Pool.east) -- (Z1-\i.west);
    \foreach \j in {1,2,80}{
      \draw[connect,white,line width=1.2] (Z1-\i.east) -- (Z2-\j.west);
      \draw[connect] (Z1-\i.east) -- (Z2-\j.west);
    }
    \draw[connect,white,line width=1.2] (Z2-\i.east) -- (Out.west);
    \draw[connect] (Z2-\i.east) -- (Out.west);
  }

  \draw[myred!40,fill=myred,fill opacity=0.02,rounded corners=2]
    (10.95,-2.35) rectangle (15.45,2.35);

  \node[align=center,mygreen!60!black] at (1,2.1) {$4$-vectors};
  \node[align=center,myblue!60!black] at (10.05,1.15) {pool};
  \node[align=center,myred!60!black] at (13.20,2.55) {output FFNN};
\end{tikzpicture}}
    \caption{Architecture of the \pn{} model used in this work, as obtained from the \lo{} hyperparameter scan (see \cref{app:hyperparameter_fit} for details). The input four-vectors are first processed by a Lorentz layer, which builds learned per-particle embeddings from invariant and equivariant features. These embeddings are then passed through two \texttt{EdgeConv} layers and mapped to the final estimator by an output \ffnn{}. The labels $(4,80)$ indicate the four final-state particles and the 80-dimensional embedding used in each \texttt{EdgeConv} layer.}
    \label{fig:particlenet_architecture}
\end{figure}

Specifically, we adopt the \pn{} architecture~\cite{Qu:2019gqs}, which represents the event as a particle cloud and relies on the \texttt{EdgeConv} layer~\cite{edgeconvref},
enhanced by a Lorentz inspired layer~\cite{Butter:2017cot}.
While ParticleNet was originally devised for high-multiplicity jet tagging, it also performs well in the more constrained setup of our study (4 final state leptons).
Our Lorentz layer is a feed-forward NN complemented by a preprocessing layer that builds Minkowski dot products and invariant mass combinations out of the raw input, making the output of that network Lorentz invariant.
Our implementation of ParticleNet then takes these outputs as the input features representing the particles and uses it as the input of the \texttt{EdgeConv} layers
following the recipe of Ref.~\cite{Qu:2019gqs} (up to the width of the intermediate layer, which has been one of the free hyperparameters in the scan).
The resulting architecture is illustrated in \cref{fig:particlenet_architecture}.

\begin{table}[tb]
    \centering
    \begin{tabular}{l|r|r|c}
                       & number of parameters & $\sigma$ \% error & reported loss \\ \hline\hline \rule{0pt}{0.8\normalbaselineskip}%
        \ffnn{}        & 4017207              & 0.08 & $1.99 \cdot 10^{-4}$ \\
        Autoencoder    & 1413985              & 0.12 & $1.86 \cdot 10^{-4}$\\
        ParticleNet    &   59861              & 0.06 & $2.96 \cdot 10^{-4}$ \\
    \end{tabular}
    \caption{Comparison on the number of parameters, the predicted cross section at \lo{} for each of the architectures and the self-reported validation loss during training. As it can be observed, the three architectures can predict the fiducial cross section with better than \% accuracy (and similarly for other distributions, with some loss of accuracy in the tails) and no particular differences can be seen during training either.}
    \label{tab:nparameters}
\end{table}

To illustrate this, in \cref{tab:nparameters} we compare the number of parameters of the three architectures considered thus far. As it can be observed, by introducing physical information into the model, we have been able to considerably reduce the number of parameters without a loss of accuracy
\footnote{The error in the fiducial cross section is close to the statistical error of the calculation, thus no hierarchy between the three architectures really exists.}.

\subsection{Random-forest regression}\label{subsec:rfr}
A random-forest regressor (\rfr{}) is also used to model the continuous label \rll{} [see \cref{eq:rLL_def_lo,eq:rLL_def}] 
as a function of the event kinematics. For the implementation, we rely on the {\tt scikit-learn} Python library \cite{Pedregosa:2011ork}.
As usual, the label is \rll{}. The input features are a subset of those in \cref{eq:input_prod_dec},
\beq\label{eq:input6}
    \pt{\PZ_1},\quad y_{\PZ_1},\quad 
    \pt{\PZ_2},\quad y_{\PZ_2},\quad
    \cos\theta^*_{\Pe^+},\quad\cos\theta^*_{\mu^+}\,.
\eeq
Compared to \cref{eq:input_prod_dec}, we have excluded the lepton-pair virtualities (dominated by the on-shell region in the considered setup), the azimuthal angles of the bosons in the laboratory frame (rather insensitive to polarisation information), and the azimuthal decay angles (sensitive to off-diagonal spin correlations, but not to diagonal ones, \ie polarisation fractions). We have also checked \emph{a posteriori}  that feeding the \rfr{} with the whole list of features in \cref{eq:input_prod_dec} does not have a sizeable impact on the results obtained with the subset in \cref{eq:input6}.

A dataset of 1.2M events was randomly split in such a way that $2/3$ of the events are used for training and the remaining $1/3$ for testing.
The \rfr{} comprises $500$ decision trees, each trained on a resampled subset of the training dataset (via boostrap). For the decision-tree construction, subsets of the input features were randomly selected at each node, split in order to reduce the correlation between trees.
No maximum depth constraint is set for the tree growth, while each terminal leaf (\emph{i.e.}\ each portion of phase space) has at least 20 events. 
Overall, such a structure leads to about 20M nodes (the total number of nodes of a RF scales with the size of the training dataset divided by the minimum event number per leaf).
While there are a few events with a true \rll{} exceeding 1 or being slightly negative, the predicted \rll{} are clipped to the physical range $[0,1]$.
The same model (which we dub $\rfr_{\rm ct}$) is applied to \lo{}, \lows{} and \nlops{} events both at the training and at the testing level. 
An uncertainty on the \rfr{} predictions is estimated through a two-stage heteroscedastic regression approach~\cite{WELSH2006860,amado2023} with a gradient-boosted decision tree from the {\tt LightGBM} library~\cite{Ke:2017hft} that has been applied to out-of-bag train residuals of the $\rfr_{\rm ct}$ model.

Although the RFR results in the next section show that the selection of features in \cref{eq:input6} are optimal to capture the differential polarisation structure of the di-boson events, we have also trained and tested an alternative model 
(which we dub $\rfr_{\rm ep}$) with the same structure as described above, but using the set of input features of \cref{eq:input17}.
A comparison between the two models is reported in \cref{app:rfr_diff_inputs}.

While tempting, a comparison between the number of parameters of a NN and a random forest is not meaningful, 
owing to the different ways the two models represent information. The number of weights and biases (trainable parameters) in a NN is fixed, given a certain architecture and is rather independent of the training dataset size. On the contrary, a proxy of the \rfr{} complexity is given by 
the total number of nodes, which typically grows linearly with the number of training events. In a sense, the \rfr{} stores information
in the structure of the decision trees itself.
Therefore, performing a one-to-one comparison of the \rfr{} complexity with the corresponding NN one is not feasible.

\section{Results} \label{sec:results}

In this section, we discuss the performance of the four different models at both the level of the fiducial cross section as well as differential distributions at various accuracies.

First, in \cref{tab:results}, the results of the four approaches for the fiducial cross sections with two longitudinally polarised bosons are shown.
The truth cross-section is obtained with the reweighting technique presented above.
The ratios \rlltilde{} of the four approaches are obtained from training datasets of different accuracy (\lo{}, \lows{}, \nlops{})
In a second step, the ratio is tested against datasets with different accuracy.
The latter datasets, which are events with unpolarised gauge bosons, can be seen as \emph{mock data} that one would access experimentally.
This procedure enables us to assess the importance of a high-accuracy dataset already at training level.

Training the four networks with a \lo{} dataset and then testing the ratio \rlltilde{} on \lo{} data provides very good results.
The various cross sections do not differ from the truth by more than half a per cent.
Doing the same with \lows{} sets leads to even better results as the maximal deviation is at the level of $0.2\%$.
On the other hand, training networks with \lo{} and testing them with \lows{} gives worse results, leading to deviations of up to $4\%$ for the \rfr{} approach.
It is worth noting that for the \autoencoder{} and \pn{} solutions, the predictions remain within half a per cent.
Finally, testing prediction with \nlops{} data and testing them with the same accuracy leads to deviations of up to a per cent for \ffnn{} and \autoencoder{} while \pn{} and \rfr{} are at the per-mille level.
Testing on \nlops{} datasets while training with \lo{} or \lows{} leads to a significant degradation, with differences of $2\%$ to $6\%$.
This reflects the importance of higher-order corrections already during the training phase, which up to now have not been considered, since they significantly modify the corresponding theoretical predictions.

\begin{table}[htb]
\begin{center}
\begin{tabular}{l|l|l|lr|lr|lr|lr}
test set & \true{} & trained with  & \multicolumn{2}{c|}{\ffnn{}} & \multicolumn{2}{c|}{\autoencoder{}} & \multicolumn{2}{c|}{\pn{}} & \multicolumn{2}{c}{\rfr{}} \\[0.3mm]
(accuracy)    & $\sigma [\fb]$ & (accuracy)    & $\sigma [\fb]$ & $\Delta [\%] $ & $\sigma [\fb]$ & $\Delta [\%] $ & $\sigma [\fb]$ & $\Delta [\%] $ & $\sigma [\fb]$ & $\Delta [\%] $\\
\hline
\hline
\multirow{1}{*}{\lo{}}    & \multirow{1}{*}{0.6567} & \lo{}    & 0.6548 & -0.3  & 0.6559 & -0.1 & 0.6565 & 0    & 0.6533(8) & -0.5 \\
\hline
\multirow{2}{*}{\lows{}}  & \multirow{2}{*}{0.6656} & \lo{}    & 0.6567 & -1.3  & 0.6686 & 0.5  & 0.6672 &  0.2 & 0.6401(1) & -3.8 \\
                          &                         & \lows{}  & 0.6655 & 0     & 0.6642 & -0.2 & 0.6651 & -0.1 & 0.6659(9) & 0  \\[1mm]
\hline
\multirow{3}{*}{\nlops{}} & \multirow{3}{*}{0.8833} & \lo{}    & 0.8490 & -3.9 & 0.8638 & -2.2 & 0.8637 & -2.2 & 0.8320(6) & -5.8  \\ 
                          &                         & \lows{}  & 0.8614 & -2.5 & 0.8595 & -2.7 & 0.8608 & -2.5 & 0.8653(1) & -2.0 \\
                          &                         & \nlops{} & 0.8711 & -1.4 & 0.8732 & -1.1 & 0.8813 & -0.2 & 0.884(1)  & -0.1
\end{tabular}

\caption{
Fiducial cross section for $\mathrm{p} \mathrm{p} \rightarrow Z_{\mathrm{L}} Z_{\mathrm{L}} \rightarrow \mathrm{e}^{+} \mathrm{e}^{-} \mu^{+} \mu^{-}$ (in $\mathrm{fb}$) in the fiducial LHC setup of \cref{eq:fidZZ}.
The first and third columns show the accuracy (\lo{}, \lows{}, \nlops{}) of the testing and the training datasets, respectively. 
The truth fiducial cross section (obtained with truth-\rll{} reweighting of \POWHEGBOX{} events) is presented in the second column.
The other columns show the fiducial cross sections for the various models and combinations of testing and training datasets. The relative differences ($\Delta$, in percentages) are given with respect to the truth.
}
\label{tab:results}
\end{center}
\end{table}

\begin{figure}[htb!]
\centering
\includegraphics[height=.30\textheight,page=1]{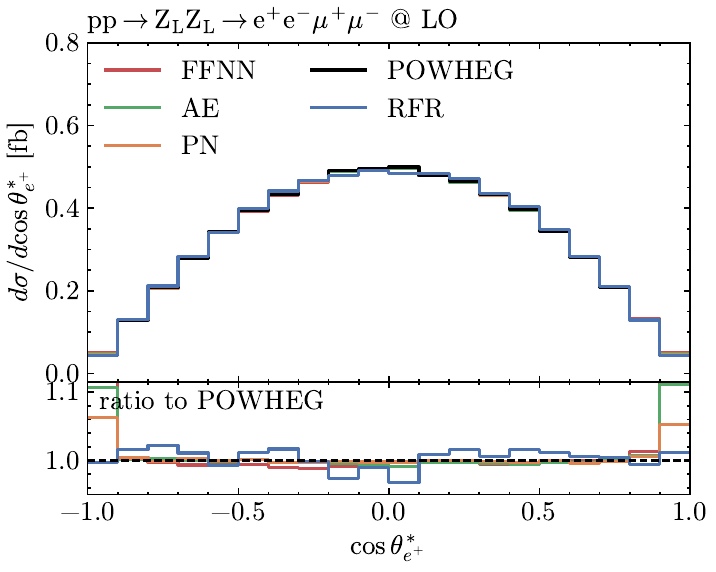} \hfill
\includegraphics[height=.30\textheight,page=2]{network_comparison_lo.pdf}
\caption{
Differential distributions in the positron decay angle in the corresponding $\PZ$-boson rest frame (left) and in the positron transverse momentum (right) for $\mathrm{p} \mathrm{p} \rightarrow \PZ_{\mathrm{L}} \PZ_{\mathrm{L}} \rightarrow \mathrm{e}^{+} \mathrm{e}^{-} \mu^{+} \mu^{-}$ (in $\mathrm{fb}$) in the fiducial LHC setup of \cref{eq:fidZZ} at \lo{} accuracy. The results of the truth-\rll{} reweighting are shown in black (dubbed \textsc{POWHEG}), while the predictions of the models are shown in green (autoencoder-like NN), red (feed-forward NN), orange (ParticleNet), and blue (random-forest regressor), respectively.
}
\label{fig:lo_costheta_ptep}
\end{figure}

Next, we turn to the differential distributions where the four approaches are compared against the true reweighting approach, named \textsc{POWHEG} in the plots.
First in \cref{fig:lo_costheta_ptep}, \lo{} predictions are compared for the positron decay angle in the corresponding $\PZ$-boson rest frame and the transverse momentum of the positron.
The decay angle shows good agreement for all approaches at the level of a couple of per cent across the whole phase space, apart from the two outer bins ($[\pm 0.9,\pm1]$).
There, the networks differ, apart from the \rfr{} approach, by $5\%$ to $15\%$.
It is worth noting that in the central region, the truth-\rll{} shape is captured slightly better by the NNs than the \rfr{}.

For the transverse momentum, all approaches agree well within a few per cent with the truth up to about $200\GeV$.
For larger energies, significant differences appear and only the \rfr{} provides a satisfactory solution.
Nonetheless, it should be noted that events with an electron transverse momentum larger than $200\GeV$ are rare, since their cross section is about four orders of magnitude lower than the bulk of the cross section.
This means that even if large differences appear in this part of phase space, they are largely negligible for phenomenological studies.

\begin{figure}[htb!]
\centering
\includegraphics[height=.30\textheight,page=1]{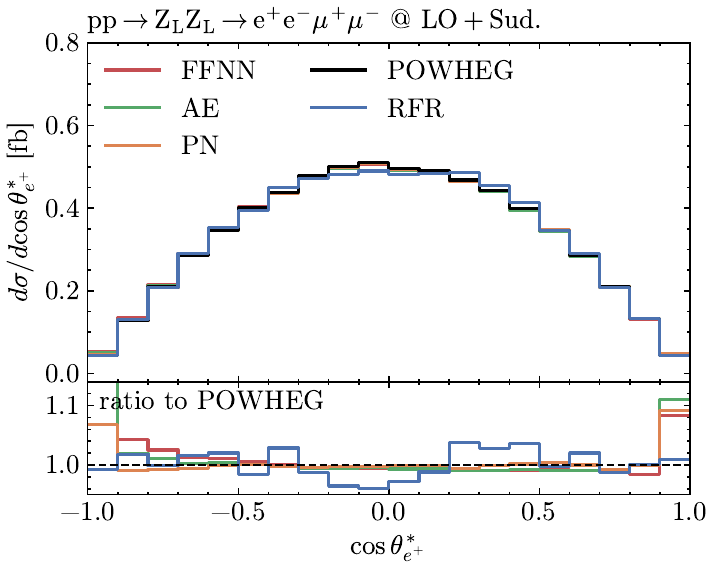} \hfill
\includegraphics[height=.30\textheight,page=2]{network_comparison_lows.pdf}
\caption{Same as \cref{fig:lo_costheta_ptep} but at \lows{} accuracy.}
\label{fig:lows_costheta_ptep}
\end{figure}

\begin{figure}[htb!]
\centering
\includegraphics[height=.30\textheight,page=1]{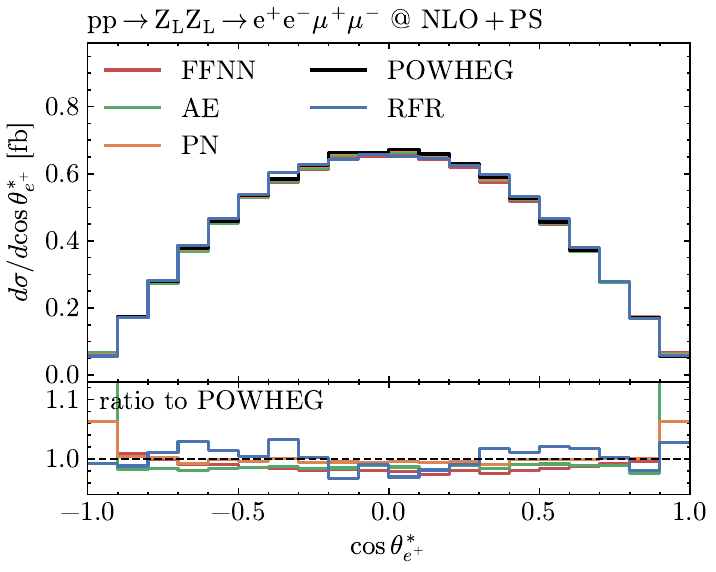} \hfill
\includegraphics[height=.30\textheight,page=3]{network_comparison_nlops.pdf}
\caption{Same as \cref{fig:lo_costheta_ptep} but at \nlops{} accuracy.}
\label{fig:nlops_costheta_ptep}
\end{figure}

Second, in \cref{fig:lows_costheta_ptep,fig:nlops_costheta_ptep}, the same observables are shown at \lows{} and \nlops{}, respectively.
In both cases, no significant differences are observed among the four approaches for the positron decay angle, except in the outer bins.
There, only the \rfr{} is catching the correct behaviour at the per-cent level while the other models differ by up to 30\% (not visible on the plots). The \pn{} is the best performing model amongst the NNs, with at most 9\% deviations in the outer bins at \lows{} and \nlops{}.
On the other hand, for the transverse momentum of the positron, the differences are not as large as in the \lo{} case.
While the \rfr{} was favoured at \lo{}, especially at large energies, it is actually providing the worst regression in this very region for \lows{} accuracy.
Instead, in this case, the \autoencoder{} and the \pn{} are providing the best results.
At \nlops{} accuracy, the situation is quite similar, with the \autoencoder{} and the \pn{} delivering the best description, while the \ffnn{} and \rfr{} predictions are deviating more from the truth-reweighting results.
Again, it should be emphasised that the prediction behaviour above $200\GeV$ is largely irrelevant for phenomenological studies and that within the physically relevant part of the phase space, the four approaches give equally good results.

\begin{figure}[htb!]
\centering
\includegraphics[height=.30\textheight,page=7]{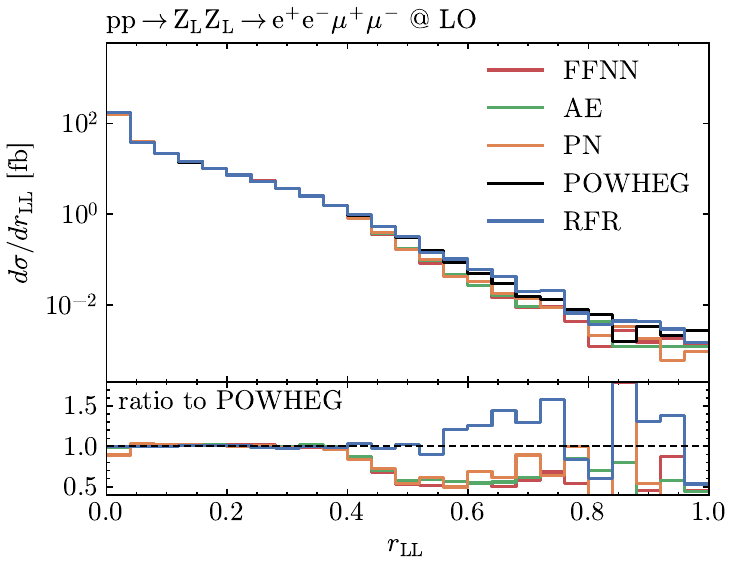} \hfill
\includegraphics[height=.30\textheight,page=9]{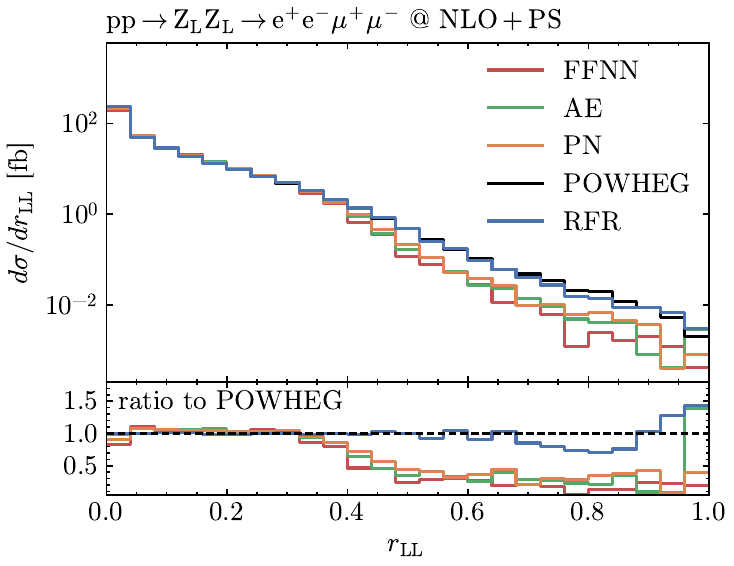}
\caption{
Distributions in the truth and predicted \rll{} weights for $\mathrm{p} \mathrm{p} \rightarrow \PZ\PZ \rightarrow \mathrm{e}^{+} \mathrm{e}^{-} \mu^{+} \mu^{-}$ (in $\mathrm{fb}$) in the fiducial LHC setup of \cref{eq:fidZZ} at \lo{} (left) and \nlops{} (right) accuracy. Same key as in \cref{fig:lo_costheta_ptep}.}
\label{fig:lo_nlops_rLL}
\end{figure}

Finally, in \cref{fig:lo_nlops_rLL}, the differential distribution in \rll{} is displayed at \lo{} and \nlops{} accuracy.
While this is not a physically observable quantity, it is of particular interest as it provides an insight into the overall goodness of the approaches over the whole phase space.
At \lo{} accuracy, all approaches provide a very good description of the reweighting procedure up to about 0.4.
Beyond this point, the approaches begin to diverge, with large statistical variation.
It is noteworthy to observe that the \rfr{} approach differs significantly from the other three.
In particular, it tends to overshoot the truth while the other three undershoot it in a rather similar fashion.
At \nlops{} accuracy, the picture is quite similar.
Once again, the \autoencoder{}, the \ffnn{} and the \pn{} perform identically.
In particular, they are very good up to $\rll{}\approx 0.35$ and are within a few per cent with respect to the truth, while above, they diverge.
Finally, only the \rfr{} performs consistently well over the full range, most likely because of the specific input choice made for the \rfr{} [see \cref{eq:input6}] which includes the decay angles in the corresponding $\PZ$-boson rest frames as features. Further details on this can be found in \cref{app:rfr_diff_inputs}.
We note that the \pn{} also has a similar input choice, including the decay angles as well, but fewer trainable parameters than the \ffnn{} or the \autoencoder{}.

Again, we should remind the reader that aiming for per cent precision in extracting the polarisation fractions means that events with a probability suppressed by two orders of magnitude are largely insignificant, and therefore large errors on them do not affect phenomenological studies.
Overall, the four approaches are satisfactory for phenomenological purposes, while the \pn{} and \rfr{} ones appear to perform best over the full range of kinematic variables. The goodness of the \pn{} model can also be appreciated in other differential observables shown in \cref{app:more_distrib}.

\section{Conclusions} \label{sec:conclusion}

Advances in particle physics proceed through the comparison of experimental data with theoretical predictions obtained from first principles.
This allows to infirm or confirm fundamental theories of elementary particle physics.
While this procedure could appear relatively simple at first sight, 
it turns out to be highly non-trivial in reality.
Indeed, ensuring that theoretical predictions are actually comparable with experimental data is usually difficult and all sorts of systematic effects, such as theoretical approximations, have to be accounted for.
Furthermore, many analyses, based on theoretical inputs, aim at extracting fundamental parameters of the theory or pseudo-observables from data.
In this endeavour, machine-learning approaches appear to be particularly useful in linking theory and experiment.

The extraction of the longitudinal polarisation of a massive gauge boson is a perfect illustration of it.
Indeed, this pseudo-observable can only be extracted from data with the help of theoretical inputs.
In Ref.~\cite{Grossi:2023fqq}, a new approach based on NNs and matrix elements was proposed.
The work was restricted to \lops{} accuracy and considered Z+j production at the LHC as a proof of concept.
In the present article, we go beyond this original work in many ways by looking at the production of polarised Z-boson pairs at the LHC.

First, we have generalised the method up to \nlops{} accuracy.
This implies defining a ratio of polarised and unpolarised amplitudes (dubbed $r_{\lambda\lambda'}$) at \nlops{} accuracy at the event level.
To that end, we have developed a reweighting procedure that acts at the level of the fully-differential NLO weights, thereby being necessarily approximate in the context of fixed-order computations matched to parton showers.
This approximation turned out to be particularly good with respect to the anticipated precision for the extraction of polarised fractions.
While we have focused on an implementation in the \POWHEGBOX~framework~\cite{Alioli:2010xd,Jezo:2015aia}, our proposal should be useful for other theoretical approaches.
As a by-product, the implementation of the reweighting procedure in the \POWHEG{} framework for unpolarised di-boson events with truth $r_{\lambda\lambda'}$ weights (with $\lambda,\lambda'=\rL,\rT$), described in \cref{subsec:rew}, is now part of the public 
{\sc VV\!\_pol}~package~\cite{Pelliccioli:2023zpd,Haisch:2025jqr}, which can be found at
\begin{center}
\url{https://gitlab.com/POWHEG-BOX/RES/User-Processes/VV_pol} .
\end{center}
It is important to point out that, while technically possible, not all polarisation combinations are equally well-suited for a reweighting. We encourage, therefore, every user to examine the reweighting approach on a case-by-case basis, as done for example in \cref{subsec:tests}.
Second, while the entire knowledge of a phase space allows to devise a reweighting procedure as highlighted above, an application to experimental events requires to account for unobserved degrees of freedom.
To that end, machine-learning approaches are particularly well-suited.
They allow for performing a regression on the ratio $r_{\lambda\lambda'}$ with an incomplete set of information, as only measurable observables can be used.
This leads to the use of a proxy for the ratio $r_{\lambda\lambda'}$, which can be used to tag the polarisation of the gauge bosons.
In the present work, we have proposed various approaches to address this:
a feed-forward NN, an autoencoder-like NN, a physics-informed NN, and a random-forest regressor.
All four approaches have been compared and discussed.
We find that all four approaches are well-suited for phenomenological purposes. However, thanks to its good performance and relatively modest number of trainable parameters, the \pn{} model stands out and is therefore our preferred choice.
To ease reproducibility and dissemination, all ML predictions are available in the public Git repository
\begin{center}
\url{https://github.com/jakoblinder/polarisation_tagging} .
\end{center}

We stress that, while we focused on a regression approach, the event-by-event predictions for \rll{} obtained in this work with various models can be used without
any modifications to perform a polarisation tagging of LHC di-boson events (simulated or real data). 
Furthermore, although the present analysis focuses on leptonic decays of electroweak bosons, the extension to hadronic decays is straightforward, provided
a \POWHEGBOX~event generator is available for the considered LHC signature.

Finally, while this work should be useful for both theoretical and experimental work aiming at exploring the polarisation of massive electroweak bosons at the LHC, we would like to point out that it also perfectly illustrates the richness
of the interplay between experimental analysis, theoretical predictions, and machine learning.

\section*{Acknowledgements}
The authors would like to thank Erik Bachmann, Uli Haisch, Mareen Hoppe, Tae Hyoun Park, Frank Siegert, and Giulia Zanderighi for fruitful discussions.
We are also grateful to Giulia Zanderighi for comments on the manuscript.
J.L.\ is especially grateful to Tae Hyoun Park for support with the GPU framework used for training the NNs.
The authors acknowledge support from the European Union (EU) COMETA COST Action (CA22130). 
J.C-M.\ acknowledges support from the Ramón y Cajal program grant RYC2023-043794-I funded by MCIN/AEI/10.13039/501100011033 and by ESF+.
G.P.\ acknowledges financial support from the EU Horizon Europe research and innovation programme under the Marie-Sk\l{}odowska Curie Action (MSCA) ``POEBLITA - POlarised Electroweak Bosons at the LHC with Improved Theoretical Accuracy'' - grant agreement Nr.~101149251 (CUP H45E2300129000).
The authors acknowledge the use of computing resources on the Max Planck Computing and Data Facility (MPCDF) in Garching to run all simulations from the resulting data was used to train the different networks, and on the Leonardo supercomputing cluster at CINECA, Italy, provided under the INFN project PML4HEP.

\appendix

\section{Additional kinematic plots}\label{app:more_distrib}

\begin{figure}[htb!]
\centering
\includegraphics[height=.30\textheight,page=3]{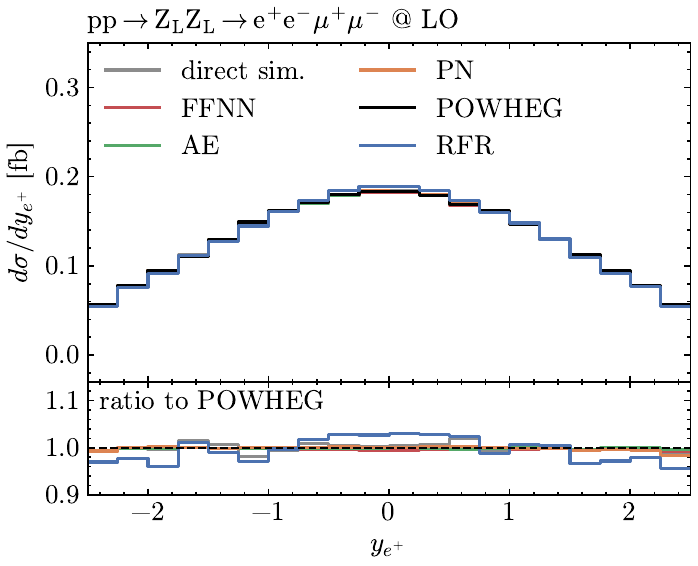} \hfill
\includegraphics[height=.30\textheight,page=5]{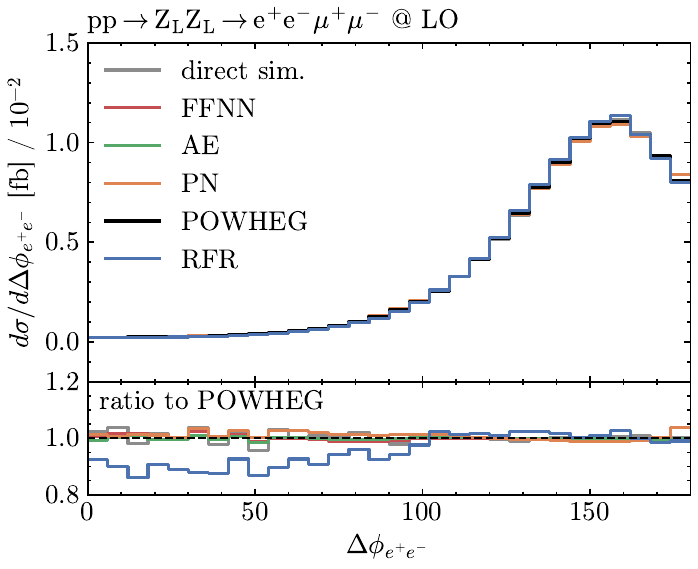}
\caption{Differential distributions in the positron rapidity (left) and in the positron--electron azimuthal separation (right) at \lo{} accuracy. Same structure as \cref{fig:lo_costheta_ptep}.}
\label{fig:lo_yep_phiepem}
\end{figure}

\begin{figure}[htb!]
\centering
\includegraphics[height=.30\textheight,page=3]{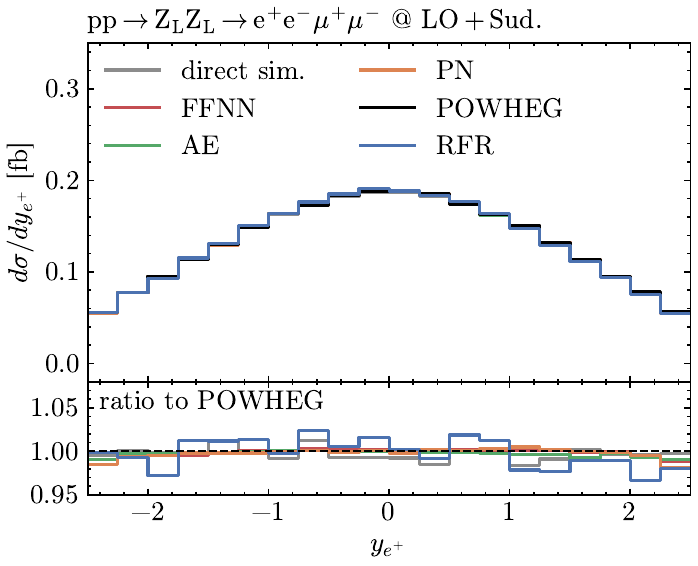} \hfill
\includegraphics[height=.30\textheight,page=5]{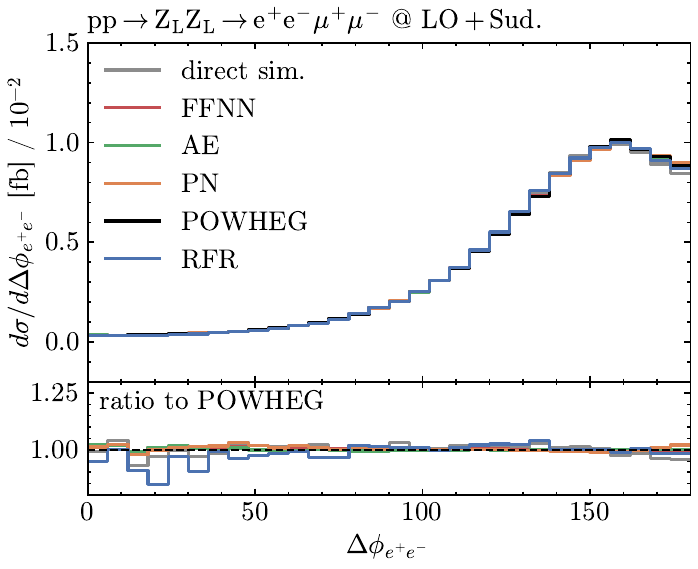}
\caption{Same as \cref{fig:lo_yep_phiepem} but at \lows{} accuracy.}
\label{fig:lows_yep_phiepem}
\end{figure}

\begin{figure}[htb!]
\centering
\includegraphics[height=.30\textheight,page=4]{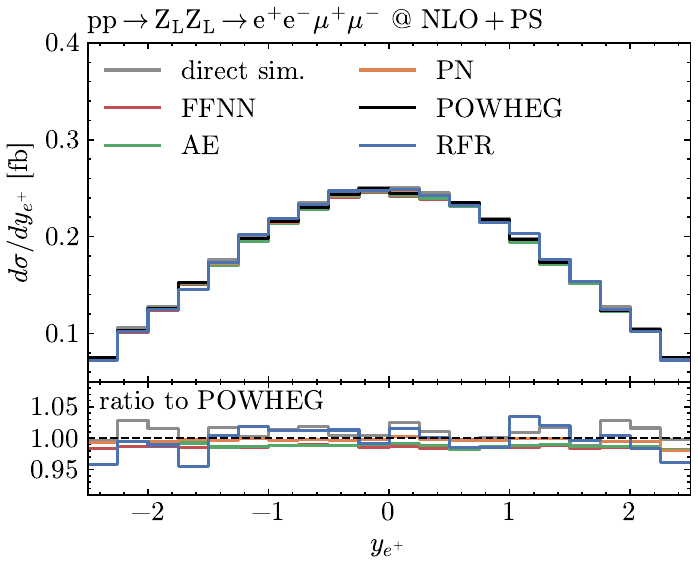} \hfill
\includegraphics[height=.30\textheight,page=6]{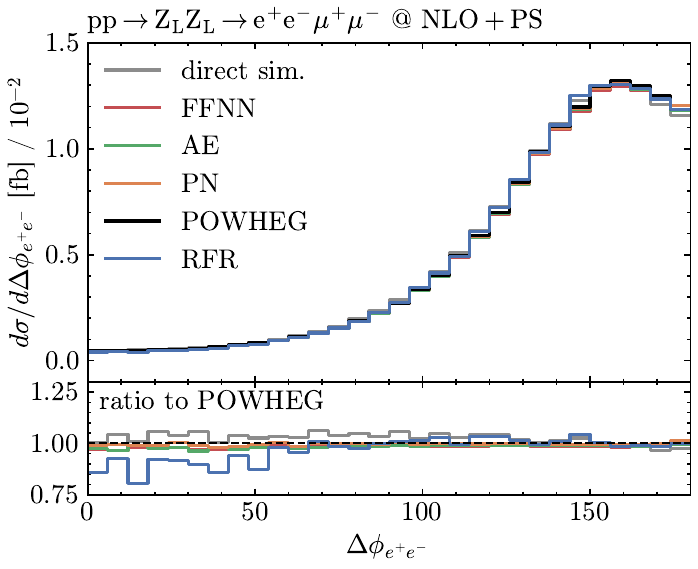}
\caption{Same as \cref{fig:lo_yep_phiepem} but at \nlops{} accuracy.}
\label{fig:nlops_yep_phiepem}
\end{figure}

\begin{figure}[htb!]
\centering
\includegraphics[height=.30\textheight,page=7]{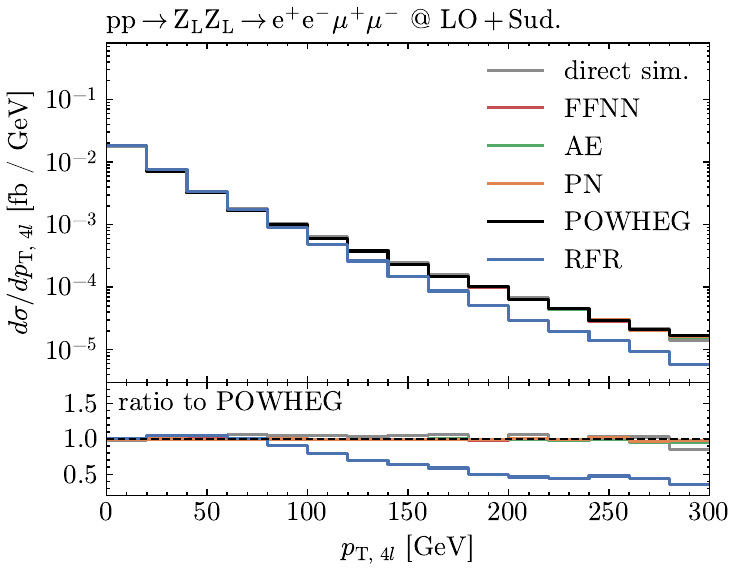} \hfill
\includegraphics[height=.30\textheight,page=8]{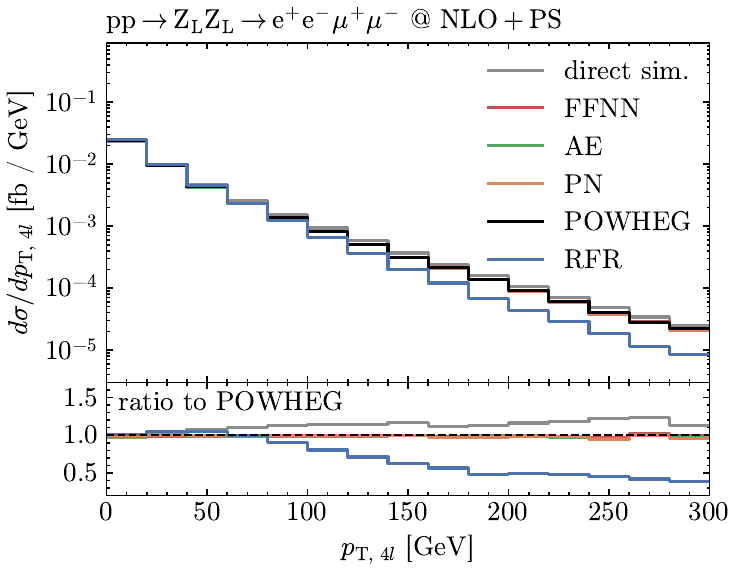}
\caption{Differential distributions in the transverse momentum of the four-lepton system at \lows{} (left) and \nlops{} accuracy. Same structure as \cref{fig:lo_costheta_ptep}. In addition, \emph{direct sim.} denotes the full polarised prediction as opposed to the \POWHEG{} reweighted prediction.}
\label{fig:lows_nlops_mepem}
\end{figure}

In this section, we provide additional differential observables for the comparisons between the four different models.
In particular, we provide them for completeness for the interested reader, but do not comment on them in detail as in the main text.
\Cref{fig:lo_yep_phiepem} provides a differential comparison for the rapidity of the positron as well as the azimuthal-angle difference between the electron and the positron at \lo{} accuracy.
\Cref{fig:lows_yep_phiepem,fig:nlops_yep_phiepem} are the same type of distributions at \lows{} and \nlops{} accuracy, respectively.
Finally, \cref{fig:lows_nlops_mepem} shows 
the transverse momentum of the four-lepton system at \lows{} and \nlops{} accuracy.
In all figures, we include the results from the direct simulation of the $\rL\rL$ signal (obtained with the \POWHEG{}~package {\sc VV\!\_pol} and dubbed \emph{direct sim.}), in addition to the four models and the reweighted \POWHEG{}~predictions.

\section{Impact of input features in \rfr{} results} \label{app:rfr_diff_inputs}

\begin{figure}[htb!]
\centering
\includegraphics[height=.30\textheight,page=1]{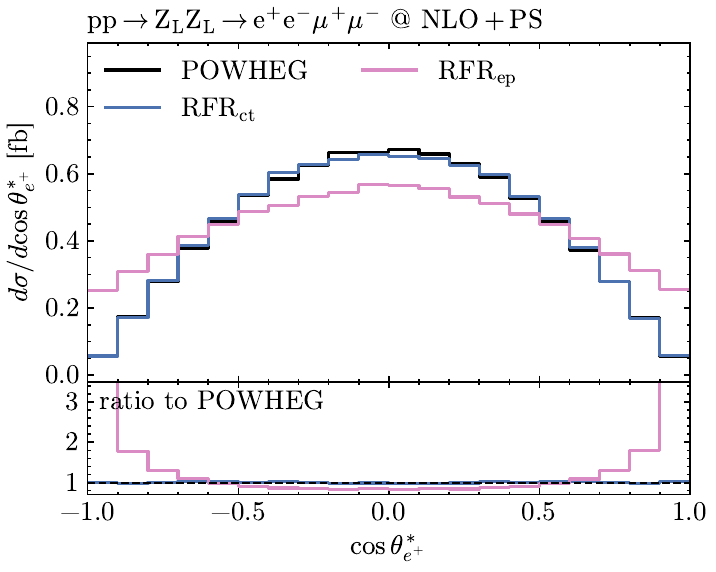} \hfill
\includegraphics[height=.30\textheight,page=3]{rfr_input_comparison_nlops.pdf}
\caption{Same $\cos\theta^*_{\rm e^{+}}$ and $
p_{\rm T, \, e^{+}}$ distributions as in \cref{fig:nlops_costheta_ptep} but showing only
results from the truth-\rll{} (black), the default \rfr{} (blue) and the alternative model $\rfr_{\rm ep}$ (pink).}
\label{fig:rfr_input_comparison}
\end{figure}

Although the \rfr{} model leads to a moderate event-by-event correlation (about 75\%), the sum of predicted weights reproduces the expected value from true \rll{} weights to less than $1\%$ accuracy, as shown in \cref{tab:results}.
In other words, while noisy across individual events, the regression is rather unbiased on the event average, therefore reproducing the fiducial LL cross-section accurately. 
The $\rfr_{\rm ct}$ model performance,  correlation and accuracy are almost identical at \lo{} and at \nlops{}, pointing out its robustness against the boost of longitudinal bosons due to the hadronic recoil.
The results in the next section show that the selection of features in \cref{eq:input6} is optimal to capture the polarisation structure of the events from decay-like observables directly connected to the single-boson polarisation state (polar decay angles) without losing information about the kinematics of the underlying $\Pp\Pp\rightarrow \PZ\PZ$ process (transverse momenta and rapidities of the two bosons).
This can be further understood by constructing an alternative model (dubbed $\rfr_{\rm ep}$) which has the very same structure as the default \rfr{} model described in \cref{subsec:rfr} but uses the variables in \cref{eq:input17} as input features. The prediction of the $\rL\rL$ cross section by this model at \nlops{} (but also at \lo{}) is just as good as the one of the default \rfr{}, reproducing at the sub-per cent level the truth-\rll{}, 
\begin{align}
\sigma^{\nlops}_{\rL\rL}(r^{\rm \rfr_{\rm ep}}_{\rL\rL}) = 0.885(2) \,{\rm fb}\,.
\end{align}
However, the differential behaviour features some deviations from the expected distributions.
The differential results are shown in \cref{fig:rfr_input_comparison}, where the results from truth-\rll{} are compared with the predictions from the two \rfr{} models ($\rfr_{\rm ct}$ and $\rfr_{\rm ep}$).
The choice of the input as in \cref{eq:input6} ($\rfr_{\rm ct}$) includes the decay angle that is considered in the left side of \cref{fig:rfr_input_comparison}, leading to a very good modelling of these observables, as shown in \cref{sec:results}.
On the contrary, since all quantities in \cref{eq:input17} are written in the laboratory frame, the decay-angle shape is poorly described by the $\rfr_{\rm ep}$ model.
This choice of input proves to be superior also in energy-dependent distributions, especially in the tails of transverse-momentum distributions. This can be observed in the right panel of \cref{fig:rfr_input_comparison}.
The default $\rfr_{\rm cp}$ reproduces the truth-\rll{} distribution in the positron transverse momentum, with at most 10\% discrepancies for $\pt{\Pe^+}\lesssim200\GeV$, while the alternative model leads to a much larger discrepancy already for  $\pt{\Pe^+}\gtrsim100\GeV$.
The comparison highlights the fact that a random forest does not have any knowledge about how observables transform under Lorentz boosts and rotations.
Being optimised to predict events as accurately as possible on average, but not to predict distributions derived from intricate transformations, the \rfr{}
benefits from either Lorentz-invariant quantities or observables written in the appropriate reference frames.

\section{Training dynamics and hyperparameter optimisation}\label{app:hyperparameters}
This \lcnamecref{app:hyperparameters} describes the training strategies used to train the NNs, described in \cref{subsec:ffnn,subsec:aenn,subsec:pn}, as well as the optimisation of their hyperparameters. Having the most interesting features, we illustrate both training and hyperparameter optimisation for the \ffnn{} at \lo{} and \lows{}
At \nlops{}, the training and hyperparameter optimisation behave qualitatively similarly to the \lo{} curves shown here, and for the \autoencoder{} network, no new features are observed either.

\subsection{Training strategy} \label{app:training_history}
\begin{figure}[htb!]
\centering
\includegraphics[height=.26\textheight,page=1]{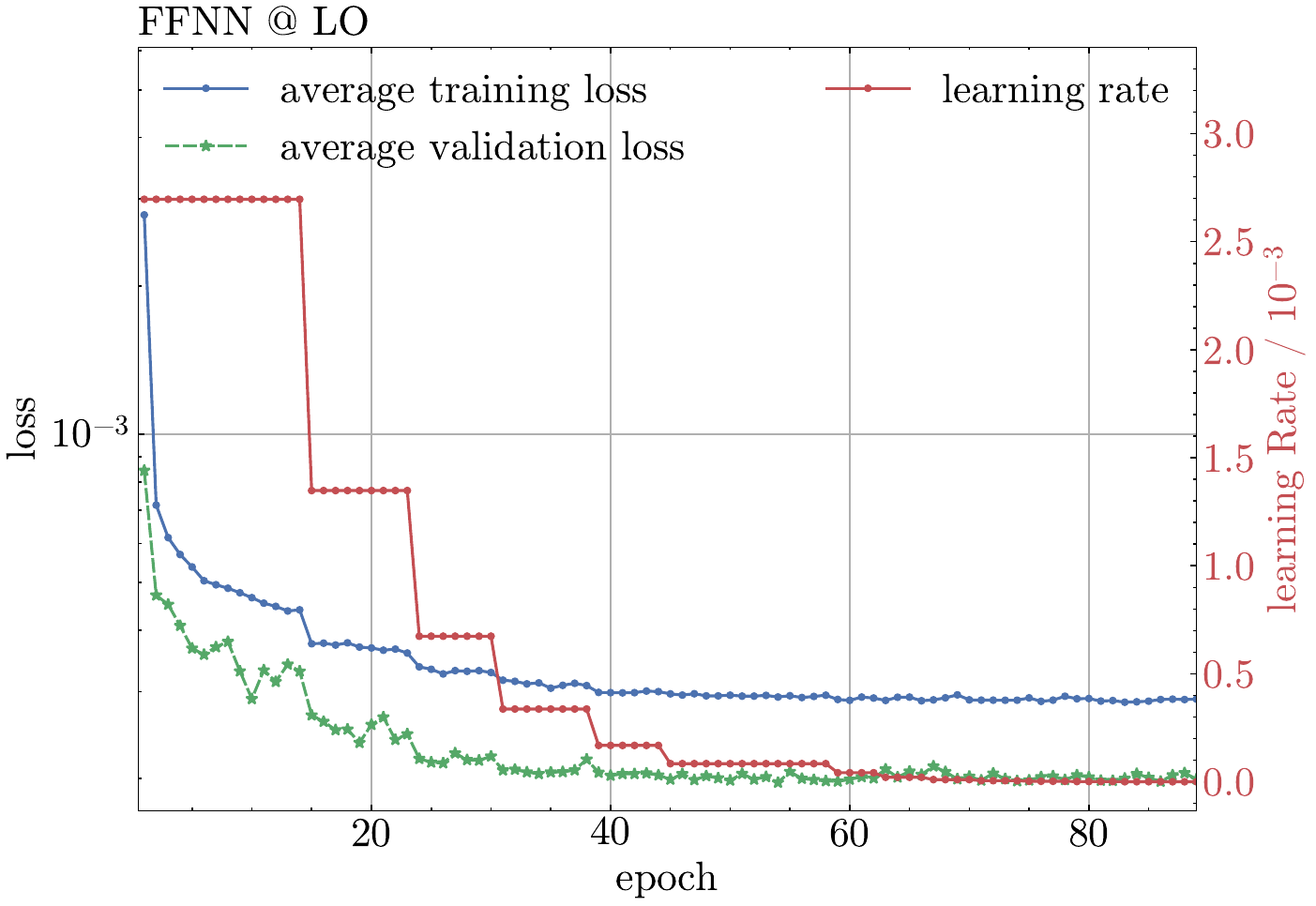} \hfill
\includegraphics[height=.26\textheight,page=1]{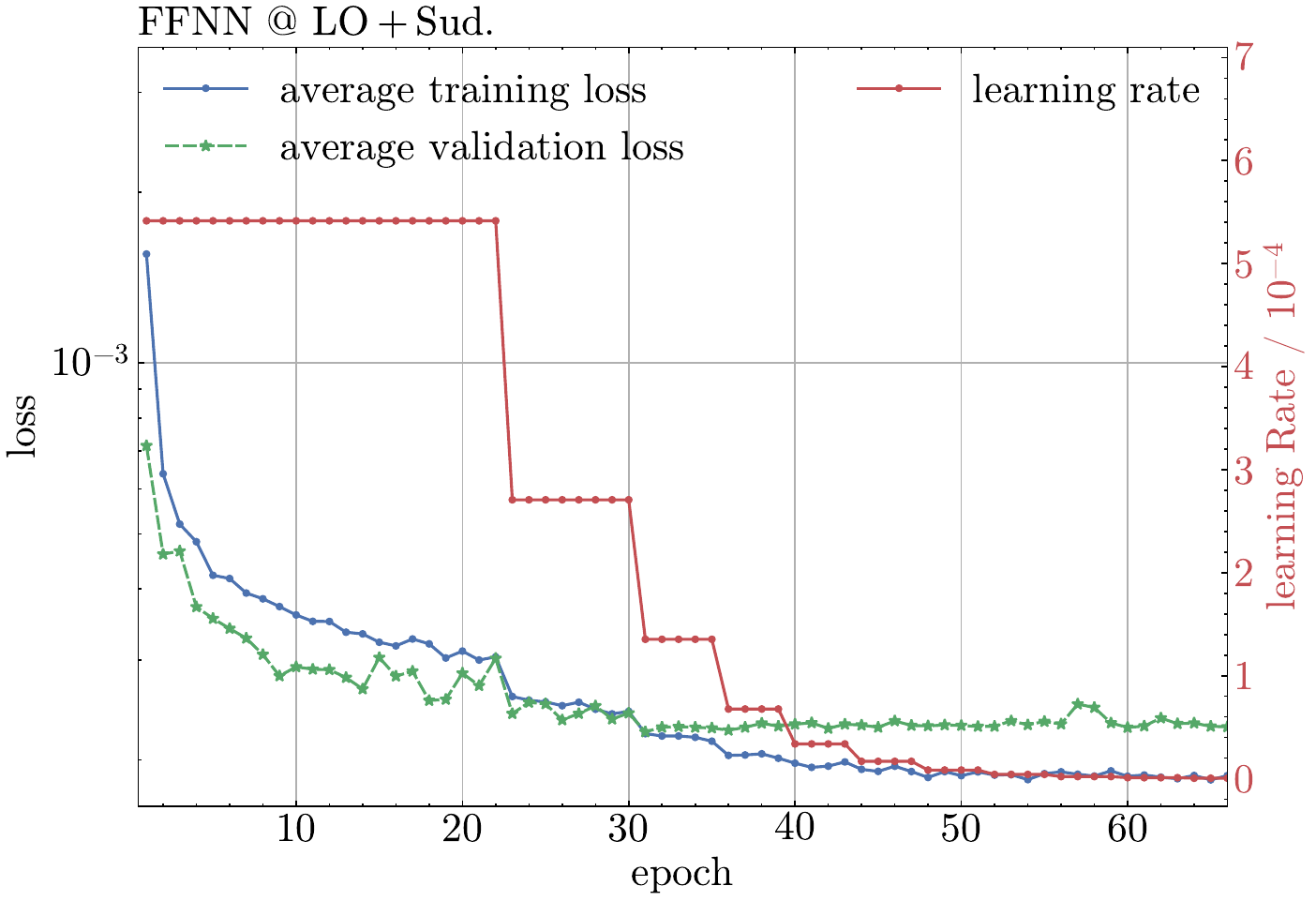}
\caption{Training history for the \ffnn{} at \lo{} (left panel) and \lows{} (right panel). The blue and green curves show the loss values for each epoch for the training and validation datasets. Their values refer to the left vertical axis. The right-hand learning rate axis illustrates the change in the learning rate across epochs.}
\label{fig:training_history}
\end{figure}

The central aim in the training of a NN is the minimisation of a loss function, for which we chose the mean squared error loss. Instead of taking, however, the ordinary mean over one batch, $b$, during the training, we take a weighted average with respect to the unpolarised cross-section:
\begin{equation}
    L =  \frac{1}{n_{b}} \sum_{b} \frac{ \sum_{b_{i}} (\rlltilde^{b_{i}} - \rll^{b_{i}})^{2} \, \sigma_{\mathrm{UU}}^{b_{i}} }{ \sum_{b_{i}} \sigma_{\mathrm{UU}}^{b_{i}}}
    \, .
    \label{eq:loss_function}
\end{equation}
This loss function ensures that phase-space regions which do not contribute to the (unpolarised) cross-section are accordingly given less importance during training.
The first sum over $b$ in \cref{eq:loss_function} runs over the different $n_{b}$ batches for each epoch, whereas $b_{i}$ denotes the different events within each batch. To avoid overfitting, the training is stopped when the validation loss is not decreasing for 35 epochs (empirically set), as can be seen, for example, in the right panel of \cref{fig:training_history}, where the validation loss is getting even bigger than the training loss, and the training is therefore stopped.

In order to avoid a stagnated learning, \ie when the training is stuck on a plateau, the learning rate is halved should there be no progress in training. The effect of this is clearly seen, for example, in the left panel of \cref{fig:training_history}, where the learning rate for each epoch is shown in red. At epoch 14, the learning rate is reduced and as a result, the training loss (blue) and the validation loss (green) decrease. The best loss value is reported for \lo{} in \cref{tab:nparameters}.

\subsection{Hyperparameter fit} \label{app:hyperparameter_fit}
The success of our learning strategies depends heavily on the hyperparameters chosen during the training, for which we are providing an overview in \cref{tab:hyperparameters}.
\begin{table}[tb]
    \centering
    \begin{tabular}{l|l}
        parameter                     & value \\ \hline
        activation function           & \relu{} \cite{glorot2011deep} \\
        batch size                    & 1000 -- 5000     \\
        learning rate                 & 0.0054 -- 0.0038 \\
        optimizer (weight decay)      & \noun{AdamW} ($10^{-4}$) \cite{loshchilov2018decoupled}  \\
        training epochs               & 66 -- 139            \\
        train--validation--test split & 60\% -- 20\% -- 20\%
    \end{tabular}
    \caption{Summary of the \ffnn{} and \autoencoder{} training hyperparameters.}
    \label{tab:hyperparameters}
\end{table}
The choice of these hyperparameters has been informed by the previous study in Ref.~\cite{Grossi:2023fqq}, motivated already in \cref{subsec:ffnn} and fix mostly the architecture of the network.
With the architecture fixed, we focus for our hyperparameter scan on the optimization hyperparameters, in particular we focus on finding the optimal batch size and learning rate, for which only a range is given in \cref{tab:hyperparameters}. In the following we detail the technical implementation of the hyperparameter scan.

\begin{figure}[htb]
\centering
\includegraphics[height=.30\textheight,page=3]{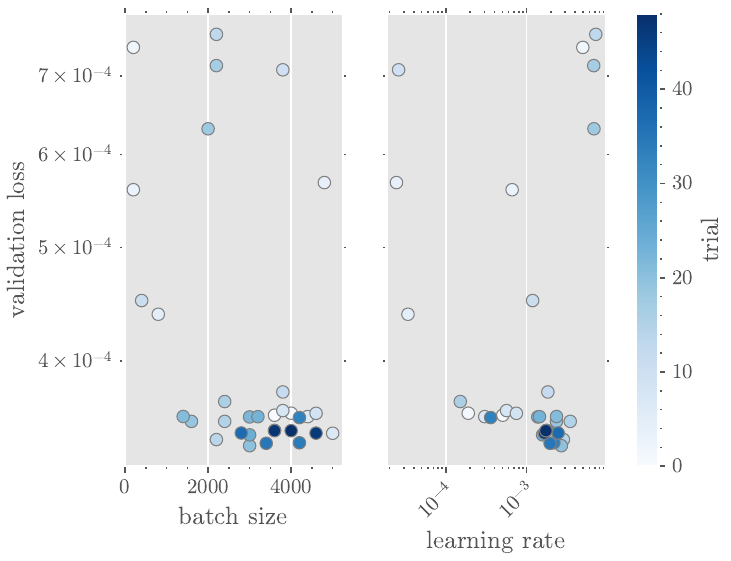} \hfill
\includegraphics[height=.30\textheight,page=3]{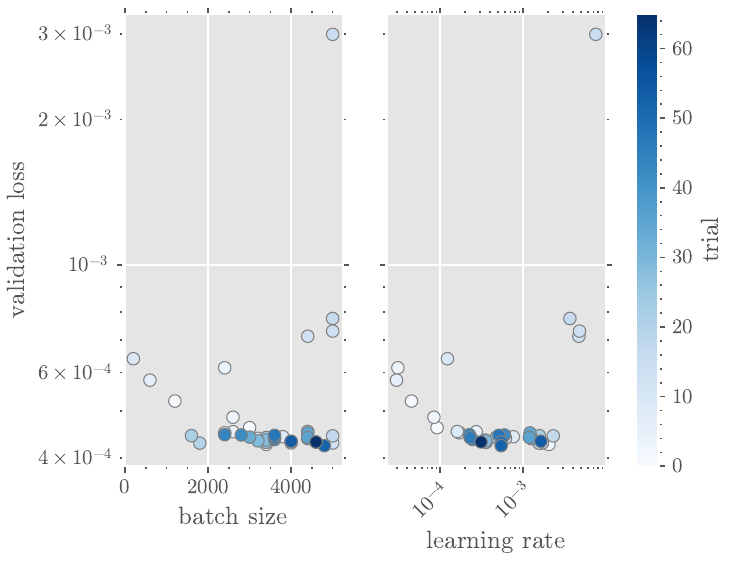}
\caption{Loss values according to \cref{eq:loss_function} for different batch sizes and learning rates for the \ffnn{} at \lo{} (left panel) and \lows{} (right panel). The colour of the dots indicates the trial number, with darker shades of blue corresponding to later trials.}
\label{fig:hyperparameter_fit}
\end{figure}

Using the \noun{Optuna}~framework~\cite{akiba2019optunanextgenerationhyperparameteroptimization}, the batch sizes and learning rates are varied in a range of $[2 \cdot 10^{-5}, 8 \cdot 10^{-3}]$ and $[200, 5000]$, respectively, resulting in networks with a minimal loss for the parameters given in \cref{tab:best_hyperparameters}. The parameters presented there are the ones used to provide the predictions in this article.
\begin{table}[htb]
    \centering
    \begin{tabular}{l|ll|ll}
                 & \multicolumn{2}{c|}{\ffnn{}} & \multicolumn{2}{c}{\autoencoder{}} \\
                 & batch size & learning rate  & batch size & learning rate         \\ \hline
        \lo{}    & 3000       & 0.0027         & 4200       & 0.0026                \\
        \lows{}  & 4800       & 0.00054        & 5000       & 0.0026                \\
        \nlops{} & 1000       & 0.0016         & 3400       & 0.0038                \\
    \end{tabular}
    \caption{Results of the hyperparameter optimisation performed with \noun{Optuna}~\cite{akiba2019optunanextgenerationhyperparameteroptimization} for the feed-forward NN and the autoencoder-like NN, described in \cref{subsec:ffnn,subsec:aenn} respectively.}
    \label{tab:best_hyperparameters}
\end{table}
Even though the values clearly differ across the perturbative orders and networks, it is the order of magnitude which matters.
This effect is illustrated in \cref{fig:hyperparameter_fit}, where the loss value with respect to a particular batch size and learning rate is shown and, except for small fluctuations, remains identical.

Finally, we note that \cref{tab:hyperparameters} also specifies a range for the number of training epochs. As explained in \cref{app:training_history}, this stems from the patience mechanism we used and is not a parameter we adjust during hyperparameter fitting.

The situation is slightly more complex for the implementation of the ParticleNet architecture in \cref{subsec:pn}, since no prior implementation exists for this problem to use as a baseline and the architecture is too different from both FNNN and AE to reutilize the same parameters.
For this hyperparameter scan, we include the width of the input and output layers of the network, a \texttt{LorentzLayer} and a two-layer perceptron, as well as the depth of the \texttt{EdgeConv} block that defines the ParticleNet architecture.
This optimization scan is performed separately at LO and NLO.
In the former case (LO) it results on a width of 80 nodes and and 2 \texttt{EdgeConv} layers within the central block, with a learning rate of $6\cdot10^{-5}$ and 1000 as batch size.
In the later (NLO) it ends up with a wider (128 nodes) and deeper (4 \texttt{EdgeConv}) network, for a total of 250901 trainable parameters, to be compared with the mere 60000 seen at LO in \cref{tab:nparameters}.
At NLO the optimization parameters are also reduced, with 256 as the batch size and a learning rate of $1.9\cdot10^{-5}$.

\FloatBarrier
\bibliographystyle{utphys.bst}
\bibliography{polvv}


\end{document}